\documentclass[a4wide,12pt]{report}
\usepackage{graphicx}

\begin{document}
\sloppy
 
 \newcommand{\hh}{{\cal H}}
\newcommand{\br}{{\bf R}}
\newcommand{\bc}{{\bf C}}
\newcommand{\bn}{{\bf N}} 

\newcommand{\rmi}{{\rm i}} 
\newcommand{\rme}{{\rm e}} 

\newcommand{\lt}{{L^2 ({\br},dx)}}
\newcommand{\ltp}{{L^2 ({\br},dp)}}
\newcommand{\hs}{{\cal H}}
\newcommand{\st}{{\cal S}(\br)}
\newcommand{\std}{{\cal S}^{\prime} (\br)}
\newcommand{\pr}{\prime}
\newcommand{\pa}{\partial}
\newcommand{\ds}{\displaystyle}
\renewcommand{\dag}{\ds{\dagger}}

\newcommand{\ri}{{\rm i}}

\newtheorem{defin}{Definition}
\newtheorem{theo}{Theorem}
\newtheorem{lem}{Lemma}
 
\hfill LYCEN 9960a  
\vskip 0.05truecm
\hfill TUW-00-06
\vskip 0.05truecm
\hfill July 2000
\vskip 0.05truecm
\hfill Revised version 

\bigskip

\thispagestyle{empty}

\begin{center}
{\bf \Huge{Mathematical surprises and Dirac's formalism}}
\end{center}
\begin{center}
{\bf \Huge{in quantum mechanics}\footnote{This work was 
supported by the Alexander von Humboldt Foundation 
while the author was on
leave of absence at the Institut f\"ur Theo\-retische Physik 
of the 
University of G\"ottingen. The revised version 
of the text was prepared 
during a sabbatical leave spent at the 
Institut f\"ur Theo\-retische Physik 
of the Technical 
University of Vienna.}}
\end{center}
\bigskip
\bigskip

\centerline{Dedicated to the memory of {\bf Tanguy Altherr}
(1963 - 1994)
\footnote{I dedicate these notes to the memory of Tanguy Altherr
who left us very unexpectedly in the mountains which he
liked so much (and where I could do some nice trips 
with him). 
As to the subject of these notes (which I had the pleasure
to discuss with him), Tanguy appreciated a lot Dirac's formalism 
and certainly knew how to apply it to real
physical problems (like the ones he worked on with an
impressive enthusiasm, energy and productivity).
Even if he did not share my preoccupations in this field,
he liked to discuss the problems related
to the general formalism of quantum physics.}}
 
\bigskip
\bigskip

\centerline{{\bf Fran\c cois Gieres}}
\bigskip
\centerline{\it Institut de Physique Nucl\'eaire}
\centerline{\it Universit\'e Claude Bernard (Lyon 1)}
\centerline{\it 43, boulevard du 11 novembre 1918}
\centerline{\it F - 69622 - Villeurbanne Cedex}
\bigskip
\bigskip

\bigskip
\begin{center}
{\bf \Large{Summary}}
\end{center}
By a series of simple examples, we illustrate 
how the lack of mathematical concern can readily lead to 
surprising mathematical contradictions in wave mechanics.
The basic mathematical notions allowing for a precise 
formulation of the theory are then summarized and it 
is shown how they lead to 
an elucidation and deeper understanding 
of the aforementioned problems. 
After stressing the equivalence between  wave mechanics 
and the other formulations of quantum mechanics, 
i.e. matrix mechanics and Dirac's abstract Hilbert space 
formulation,  we devote the second part of our paper 
to the latter approach: 
we discuss the problems and shortcomings of this 
formalism as well as those of the bra and ket notation 
introduced by Dirac in this context. In conclusion, 
we indicate how all of these problems can be solved 
or at least avoided.

\newpage
 
\tableofcontents
 
\newpage
 
\setcounter{page}{1}

\chapter{Introduction}

Let us first provide an overview of the present article. 
The first part,  covering {\em section 2}
and some complements gathered in the  
{\em appendix}, 
deals with mathematical surprises in quantum theory. It 
is accessible to all 
readers who have some basic notions in wave mechanics. 
The second part, consisting of {\em section 4},   
 deals with Dirac's abstract Hilbert space approach and 
his bra and ket notation; it 
is devoted to a critical study 
of this formalism which has become the  
standard mathematical language 
of quantum physics during the last decades. 
Both parts are related by  {\em section 3} 
which discusses the equivalence between wave mechanics
and the other formulations of quantum mechanics.  
 As a consequence of this equivalence, 
the mentioned surprises  also manifest themselves in 
these other formulations and in particular 
in Dirac's abstract Hilbert space approach. 
In fact, one of the points
we want to make in the second part, 
is that the lack of mathematical concern 
which is practically inherent in Dirac's symbolic calculus,
is a potential source for surprises and that it is harder, 
if not impossible, to
elucidate unambiguously all mathematical contradictions within this
formalism. 

Let us now come to the motivation for our work and 
present in greater detail the outline of our discussion. 

\bigskip 

\noindent {\bf Mathematical surprises}

\medskip 

In the formulation of physical theories, 
a lack of mathematical concern can 
often and readily lead to apparent
contradictions which are sometimes quite 
astonishing. This is particularly true for quantum mechanics 
and we will illustrate this fact by 
a series of simple examples to be presented in {\em section 2.1}.   
In the literature, such contradictions appeared in the study 
of more complicated physical phenomena and they even brought into 
question certain physical effects like the Aharonov-Bohm
effect \cite{ab}.
These contradictions can only
be discarded by appealing to a more careful mathematical
formulation of the problems, a formulation which
often provides a deeper physical understanding
of the phenomena under investigation.
Therefore, we strongly encourage the reader to look himself 
for a solution 
of all problems, eventually after a reading of  
{\em section 2.2} in which we review the 
the mathematical formalism of wave mechanics. 
In fact, the latter review also mentions some 
mathematical tools 
which are not advocated in the majority of
 quantum mechanics textbooks, though they are 
well known in mathematical 
 physics. 
The detailed solution of all of the raised problems
 is presented in the {\em appendix}.
Altogether, this discussion 
provides an overview of the subtleties 
of the subject and of the tools for handling them 
in an efficient manner.

\bigskip 

\noindent {\bf Dirac's formalism}

\medskip 

Since all of our examples of mathematical surprises 
are formulated in terms of  
the language of wave mechanics, one may wonder whether 
they are also present in other formulations of quantum theory.
In this respect, we recall that 
there are essentially three
different formulations or `representations' that are used 
in quantum mechanics for the description of the states of
a particle (or of a system of particles): wave mechanics,
matrix mechanics and the invariant formalism.
The first two rely on concrete Hilbert spaces, the
last one on an abstract Hilbert space. In general,
the latter formulation is presented using the
bra and ket notation that  Dirac developed from  
1939 on, and that he introduced 
in the third edition of his celebrated 
textbook on the principles of quantum mechanics
\cite{d}.
 Let us briefly
recall its main
ingredients which will be discussed in the main body of the 
text:
\begin{eqnarray*}
& \bullet & \quad
| \Psi \rangle \quad {\rm and} \quad \langle \Psi |
\\
& \bullet & \quad
A | \Psi \rangle \quad {\rm and} \quad \langle \Psi |A^{\dag}
\\
& \bullet & \quad
\langle \Phi |A|\Psi  \rangle
\\
& \bullet & \quad
\left\{ |n \rangle \right\}_{n \in \bn }
\quad {\rm and} \quad
\left\{ |x \rangle \right\}_{x \in \br }
\\
& \bullet & \quad
|n_1, n_2,... \rangle
\ \mbox{associated with a complete 
system of commuting observables} 
\ \left\{ A_1, A_2 ,... \right\}
.
\end{eqnarray*}
This notation usually goes together
with a specific interpretation of mathematical
operations given by Dirac.

In this respect, it may be worthwhile to mention that 
Dirac's classic monograph \cite{d} (and thereby the majority of 
modern texts inspired by it) contains a fair number of 
statements which are ambiguous or incorrect from the 
mathematical point of view: these points have been 
raised and discussed by J.M.Jauch \cite{jau}.
The state of affairs can be described as follows \cite{grau}: 
``Unfortunately, the elegance, outward clarity and strength of Dirac's
formalism  are  gained at the expense of introducing 
mathematical fictions. [...] One has a formal 
`machinery' 
whose significance is impenetrable, especially for the beginner, 
and whose problematics cannot be recognized by him." 
Thus, the verdict of major mathematicians like J.Dieudonn\'e
is devastating \cite{dieu}:
``When one gets to the mathematical theories which are at the 
basis of quantum mechanics, one realizes that the attitude of certain 
physicists in the handling of these theories truly 
borders on the delirium. [...] 
One has to wonder what remains in the mind of a student 
who has absorbed this unbelievable accumulation of nonsense,
a real gibberish! It should be to believe that today's 
physicists are only at ease in the vagueness, the obscure and 
the contradictory."
Certainly, we can blame many mathematicians for their 
intransigence and for their refusal to make the slightest 
effort to understand statements which lack rigor, 
 even so their judgment should give us something to think about.
By the present work,   
we hope to contribute in a constructive 
way to this  reflection.

In {\em section 3}, we will define Dirac's notation
more precisely while recalling some mathematical facts 
and evoking the historical development of 
quantum mechanics. 
In {\em section 4}, we will successively discuss
the following questions:
 
\begin{enumerate}
\item
 
Is any of the three representations to be preferred to the other
ones from the mathematical or practical point of view?
In particular, we discuss the status of the {\em invariant formalism}
to which the preference is given in the majority of recent
textbooks.
\item
What are the advantages, inconveniences and problems
of {\em Dirac's notations} and of their interpretation?
(The computational rules inferred from these notations and 
their interpretation are usually applied in the framework 
of the invariant formalism and then define a symbolic 
calculus - as emphasized by Dirac himself 
in the introduction of his classic text \cite{d}.) 

\end{enumerate}
 
To anticipate our answer to these questions, we already indicate
that we
will reach the conclusion that the 
{\underline{\em systematic}} application
of the invariant formalism and the {\underline{\em rigid}} 
use of Dirac's
notation - which are advocated in the majority of modern treatises
of quantum mechanics -
are neither to be recommended from a mathematical nor
from a practical point of view.
Compromises which retain the advantages of these
formalisms while avoiding their shortcomings will be
indicated. The conclusions which can be drawn for the 
practice and for the teaching of quantum theory 
are summarized in the {\em final section} 
(which also includes a short guide to the literature).

\chapter{Wave mechanics}

\section{Mathematical surprises in quantum mechanics}

Examples which are simple from the mathematical point of view
are to be followed by examples which are more sophisticated 
and more interesting from the physical point of view.
 All of them are formulated within the framework
of wave mechanics and in terms of the standard mathematical language of quantum
mechanics textbooks.
The theory of wave mechanics being equivalent to the
other formulations of quantum mechanics, the 
problems we mention 
are also present in the other formulations,
though they may be less apparent there.
The solution of all of the raised problems will be implicit
in the subsequent  
section in which we review the mathematical formalism 
of wave mechanics. It is spelled out in detail
in the appendix
(but first think about it for yourself before looking 
it up!).

\bigskip
 
\noindent 
{\bf Example 1}
 
\medskip
 
For a particle in one dimension, the operators of
momentum $P$ and position $Q$ satisfy Heisenberg's canonical
commutation relation
\begin{equation}
\label{hei} 
[ P, Q ] = \ds{\hbar \over \ri} \, {\bf 1}
\ \ .
\end{equation}
By taking the trace of this relation, one finds a vanishing
result for the left-hand side,
${\rm Tr} \, [P,Q]=0$, whereas
${\rm Tr} \,
( \ds{\hbar \over \ri} \, {\bf 1} ) \neq 0$. What is the conclusion?
 
\bigskip

\noindent 
{\bf Example 2}
 
\medskip

Consider wave functions
$\varphi$ et $\psi$ which are square integrable on $\br$
and the momentum operator $P=
\ds{\hbar \over \ri} \, \ds{d \over dx}$.
Integration by parts yields\footnote{The complex conjugate
of $z \in {\bf C}$ is denoted by $\bar z$ or $z^{\ast}$.} 
\[
\int_{-\infty}^{+\infty} dx \,  \overline{\varphi (x)} \,
(P \psi ) (x ) \ = \
\int_{-\infty}^{+\infty} dx \,
\overline{(P\varphi )(x)} \, \psi (x)   \, + \,
\ds{\hbar \over \ri} \, \left[ \left(
\overline{\varphi} \, \psi \right) (x) \right] _{-\infty}^{+\infty}
\ \ .
\]
Since $\varphi$ and $\psi$ are square integrable, one usually concludes
that these functions
vanish for $x \to \pm \infty$. Thus, the last term in the previous 
equation vanishes, which implies that the operator $P$
is Hermitian. 

However, the  textbooks of mathematics  tell us
that square integrable functions do, in general, not admit a
limit  for $x \to \pm \infty$ and therefore they do not necessarily 
vanish at infinity. 
There are even functions which are continuous and 
square summable
on $\br$ without being bounded at infinity
\cite{ri}: an example of such a function is given by
$f(x) = x^2 \, {\rm exp}\, (-x^8\, {\rm sin}^2 \, x)$,  
cf. figure 
2.1 where the period of the function has been multiplied 
by a factor $20$ in order to increase the number of 
oscillations. 

\bigskip 

\begin{figure}[h!]
\centerline{\includegraphics*[height=4cm,angle=0]{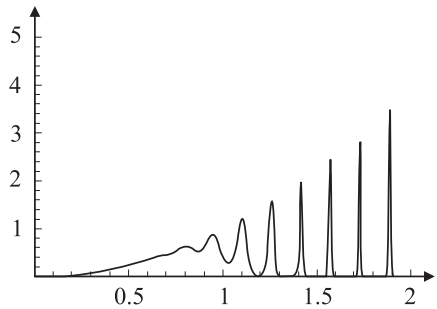}}
\caption{Graph of $f(x) = x^2 \, 
{\rm exp}\, [-x^8\, {\rm sin}^2 \, (20x)]$}
\end{figure}

\bigskip

\noindent We note that this example 
essentially 
amounts to a refinement of the following better known example
\cite{go} of a function which 
is continuous, positive and integrable on $\br$, 
though it  does not tend to zero for
$x \to \pm \infty$ (cf. figure 2.2): consider the ``shrinking
comb function''
$f(x) = \sum_{n=2}^{\infty} f_n(x)$ where $f_n$ vanishes on
$\br$, except on an interval of width $\ds{2 \over n^2}$
centered at $n$, where the graph of $f_n$ is a triangle, which is
symmetrical with respect to $n$ and of height $1$. 

\bigskip 

\begin{figure}[h!]
\centerline{\includegraphics*[height=3cm,angle=0]{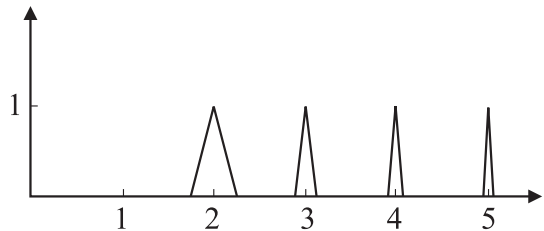}}
\caption{Shrinking comb function}
\end{figure}

\bigskip

\noindent 
The area
of this triangle being $\ds{1 \over n^2}$,
we have
\[
\int_{-\infty}^{+\infty} dx \, f(x)  =
\sum_{n=2}^{\infty} \, \ds{1 \over n^2} \, < \infty
\ \ ,
\]
but the function
$f$ does not tend to zero for $x \to +\infty$.

\noindent 
Can one conclude that the operator $P$ is Hermitian
in spite of these facts and if so, why?
 
\bigskip

\noindent 
{\bf Example 3}
 
\medskip

Consider the operators
$P= \ds{\hbar \over \ri} \, \ds{d \over dx}$ and `$Q=$ multiplication by
$x$' acting on wave functions depending on
$x \in \br$.
Since $P$ and $Q$ are Hermitian operators, the operator
$A =  PQ^3 +Q^3 P$
also has this property, because its adjoint is given by
\[
A^{\dag} = ( PQ^3 +Q^3 P )^{\dag} =
Q^3 P + PQ^3 = A
\ \ .
\]
It follows that all eigenvalues of $A$ are real.
Nevertheless, one easily verifies that
\begin{equation}
\label{fon}
Af = \ds{\hbar \over \ri} \, f
\qquad {\rm with} \quad
f(x) \ =
\left\{
\begin{array}{ll}
\ds{1\over \sqrt{2}} \, |x|^{-3/2} \; {\rm exp} \,
\left( {-1 \over 4x^2} \right)
& {\rm for} \ x \neq 0
\\
0
& {\rm for} \ x = 0 \ \ ,
\end{array}
\right.
\end{equation}
which means that $A$ admits the complex eigenvalue
$\hbar / \ri$.
Note that the function $f$ is infinitely differentiable on
$\br$ and that it is square integrable, since
\[
\int_{-\infty}^{\infty} dx \, |f(x)|^2 \, = \, 2
\int_0^{\infty} dx \, |f(x)|^2 \, = \,
\int_0^{\infty} dx \, x^{-3} e^{-1/(2x^2)} \, = \,
\left[
e^{-1/(2x^2)} \right]_0^{\infty} \, = \, 1
\ \ .
\]
Where is the error?
 
\bigskip

\noindent 
{\bf Example 4}
 
\medskip

Let us consider a particle confined to the interval
$[0,1]$ and described by a wave function $\psi$
satisfying the boundary conditions $\psi(0) =0=\psi(1)$.
Then the momentum operator $P=\ds{\hbar \over \ri} \,
\ds{d \over dx}$ is Hermitian, since the surface term
appearing upon integration by parts vanishes:
\begin{equation}
\label{itg}
\int_0^1 dx \, \left( \overline{\varphi} \,
(P\psi) -
(\overline{P\varphi} ) \, \psi \right) (x)
=
\ds{\hbar \over \ri} \, \left[ \left(
\overline{\varphi} \, \psi \right) (x) \right] _{0}^{1} = 0
\ \ .
\end{equation}
Since $P$ is Hermitian, its eigenvalues are real.
In order to determine the latter, we note that the
eigenvalue equation,
\[
(P \psi_p)(x) = p \,  \psi_p(x)
\qquad (p \in {\bf R} \ , \ \psi_p \not \equiv 0)
\ \ ,
\]
is solved by
$\psi_p (x)= c_p \, {\rm exp} \, ( \ds{\ri \over \hbar} px )$ 
with $c_p \in \bc - \{ 0 \}$.
The boundary condition $\psi_p (0) =0$ now implies
$\psi_p \equiv 0$, therefore $P$ does not admit any eigenvalues.
Nevertheless, the spectrum
of $P$ is the entire complex plane  \cite{sg}
and $P$ does not represent an observable.
How can one understand these results
which seem astonishing?

\bigskip
 
\noindent 
{\bf Example 5}
 
\medskip

 If one  introduces polar coordinates in the plane or spherical 
coordinates in space, then the polar angle $\varphi$ 
and the  component $L_z$ of angular momentum 
are canonically conjugate variables in classical mechanics. 
In quantum theory, the variable $\varphi$ becomes the operator 
of `multiplication of the wave function $\psi (\varphi)$ by $\varphi$' 
and $L_z = \ds{\hbar \over \ri} \, \ds{\pa \over \pa \varphi}$,
which implies the commutation relation 
\begin{equation}
\label{lzp}
[ L_z ,  \varphi ] = {\hbar \over \ri} \, {\bf 1}
\ \ .
\end{equation}
These operators acting on periodic wave functions  
(i.e. $\psi (0) = \psi (2\pi)$) are Hermitian. 
Furthermore, $L_z $ admits a complete system of orthonormal 
eigenfunctions $\psi_m$,
\begin{equation}
L_z \psi_m =  m \hbar \, \psi_m
\qquad {\rm with} \quad
\psi_m (\varphi ) = {1 \over \sqrt{2 \pi}} \;
{\rm exp} \, (\ri
m \varphi )
\quad {\rm and} \quad
m \in {\bf Z}
\ \ .
\end{equation}
(For the wave functions $\psi$, we only specify the dependence on the 
angular variable $\varphi$ and for the orthonormalisation, we 
refer to the standard scalar product for square integrable functions 
on the interval $[0, 2\pi )$ :
\[
\langle \psi_1 , \psi_2 \rangle = \int_0^{2 \pi}
d\varphi \ \overline{\psi_1 (\varphi )} \, \psi_2 (\varphi )
\ \ .)
\]
By evaluating the average value of the operator 
$[ L_z ,  \varphi ]$ in the state $\psi_m$ \cite{car,grau}
and by taking into account the fact that 
$L_z $ is Hermitian, one finds that 
\begin{eqnarray}
{\hbar \over \ri} \ = \
\langle \psi_m ,
{\hbar \over \ri} \, {\bf 1} \, \psi_m \rangle
& \stackrel{(\ref{lzp})}{=} &
\langle \psi_m , L_z  \varphi \, \psi_m \rangle -
\langle \psi_m , \varphi L_z \, \psi_m \rangle
\nonumber \\
& = &
\langle L_z ^{\dag} \, \psi_m , \varphi \, \psi_m \rangle -m \hbar
\, \langle \psi_m , \varphi \, \psi_m \rangle
\label{deriv}
\\
& = &
( m \hbar -  m \hbar )
\, \langle \psi_m , \varphi \, \psi_m \rangle
\ = \ 0
\ \ .
\nonumber
\end{eqnarray}
There must be a slight problem somewhere... 
(We note that similar problems appear for the phase 
and number operators which are of interest 
in quantum optics \cite{car,qo}.)

\bigskip

\noindent 
{\bf Example 6}
 
\medskip

Let us add a bit to the confusion of the previous example!
In 1927, Pauli noted that the 
canonical commutation relation (\ref{hei}) implies 
Heisenberg's uncertainty relation 
$\Delta P \cdot \Delta Q  \geq \ds{\hbar \over 2}$
by virtue of the Cauchy-Schwarz inequality.
Since the commutation relation (\ref{lzp}) has the same form as 
(\ref{hei}), one can derive, in the same way, the 
uncertainty relation 
\begin{equation}
\label{inc}
\Delta L_z \cdot \Delta \varphi \geq \ds{\hbar \over 2}
\ \ .
\end{equation}
The following physical reasoning shows that this inequality 
cannot be correct \cite{ju, car, gap}.
One can always find a state for which 
$\Delta L_z  < \hbar /  4\pi$
and then the uncertainty for the angle $\varphi$ has to be larger 
than $2 \pi$, which does not have any physical sense, since 
$\varphi$ takes values in the interval $[0,2 \pi )$.
How is it possible that relation (\ref{lzp}) is correct,
though the conclusion (\ref{inc}) is not?  

By the way, this example shows that the 
uncertainty relation 
$\Delta A\cdot \Delta B  \geq \ds{1 \over 2} \,
| \, \langle [ A,B] \rangle \, |$ 
for any two observables $A$ and $B$ (whose derivation can be found 
in most quantum mechanics texts) is not valid in such a generality.

\bigskip
 
\noindent 
{\bf Example 7}
 
\medskip

Let us consider a particle of mass $m$ in the infinite potential well  
\[
V(x) = \left\{
\begin{array}{ll}
0
& {\rm if} \ \; |x| \leq a \quad  \ (a >0)
\\
\infty
& {\rm otherwise} \ .
\end{array}
\right.
\]
The Hamiltonian for the particle confined to the inside of the well 
is simply 
$H = \ds{-\hbar^2 \over 2m}
\ds{d^2 \over dx^2}$.
Let
\begin{equation}
\label{psi}
\psi(x) =
\ds{\sqrt{15} \over 4 a^{5/2}} \, (a^2 - x^2) \qquad
{\rm for} \ |x| \leq a \quad  (\, {\rm and} \ \, \psi (x) =0 \ \,
{\rm otherwise} \, )
\end{equation}
be the normalized wave function of the particle at a given time. 
Since $H^2 \psi =
\ds{\hbar^4 \over 4 m^2} \, \ds{d^4 \psi \over dx^4} =0$,
the average value of the operator 
$H^2$ in the state $\psi$
vanishes :
\begin{equation}
\label{caf}
\langle H^2 \rangle_{\psi} = \langle \psi, H^2 \psi \rangle
= \int_{-a}^{+a} dx \; \overline{\psi (x)} \, (H^2  \psi)(x) = 0
\ \ .
\end{equation}
This average value can also be determined from the eigenvalues and 
eigenfunctions of $H$,
\begin{equation}
\label{pui}
H \varphi_n = E_n \varphi_n
\qquad {\rm with} \qquad
E_n = \ds{\pi^2 \hbar^2 \over 8ma^2} \, n^2
\qquad (n=1,2,...)
\ \ ,
\end{equation}
by applying the formula 
\begin{equation}
\langle H^2 \rangle_{\psi} = \sum_{n=1}^{\infty} E_n^2 \, p_n
\qquad {\rm with} \ \, p_n = | \langle \varphi_n ,\psi \rangle | ^2
\ \ .
\end{equation}
Proceeding in this way, 
one definitely does not find a vanishing result, because  
$E_n^2 >0$ and $0\leq p_n \leq 1, \, \, \sum_{n=1}^{\infty} p_n =1$.
In fact, the calculation yields 
$\langle H^2 \rangle_{\psi}
= \ds{15\hbar^4 \over 8m^2 a^4}$. 
Which one of these two results is correct and where does the 
inconsistency come from? \cite{grau}

\section{Clues to the understanding: the mathematical formalism}

The problems and contradictions presented in the previous 
subsection can only be elucidated by resorting to a more 
precise mathematical formulation. Therefore, we now
summarize 
the basic mathematical tools \cite{ri,sg}\cite{af}-\cite{krey}
of wave mechanics in three
subsections. 
The first two are sufficient 
for  tackling the raised problems (and actually 
provide already some 
partial solutions); the last one 
has been added in order 
to round up the discussion and to provide
a solid foundation for the subsequent section 4.  

We consider the motion of a particle on a straight line parametrized by 
$x \in {\bf R}$. The generalization to a bounded interval, to three
dimensions, to the spin or to a system of particles does not present
any problems.

\subsection{The space of states}
 
Born's probabilistic interpretation  
of the wave function $\psi$
requires that 
\[
\int _{\br} | \psi (x,t)|^2 dx =1 
\qquad {\rm for \ \; all} \ \ t \in \br 
\ \ .
\]
Thus, the wave function which depends continuously on 
the time parameter $t$, must be  
square integrable with respect to the space variable $x$:
\begin{eqnarray*}
\psi & : & {\bf R} \times {\bf R} \ \to \ {\bf C}
\\
& & \ (x, t) \ \ \, \mapsto \ \psi (x,t) \equiv \psi_t (x)
\qquad {\rm with} \quad \psi_t \in L ^2 ({\bf R} , dx )
\ \ .
\end{eqnarray*}
Here, $\lt$ is 
the space of square integrable functions,  
$$
\lt = \{ f: {\bf R} \to {\bf C} \ \mid \ \int_{{\bf R}} dx \,
| f(x) |^2 < \infty \}
\ \ ,
$$
with the scalar product
\begin{equation}
\label{ip}
\langle
f, g \rangle _{L^2} = \int_{{\bf R}} dx \; \overline{f(x)} \, g(x)
\qquad \mbox{for} \ \; f,g \in \lt
\ \ .
\end{equation}
For later reference, we recall that 
this space is related by the Fourier transformation to the Hilbert
space 
$L^2 ({\bf R} , dp )$ of wave functions depending on the momentum  $p$ :
\begin{eqnarray}
\label{f}
{\cal F} \ : & \lt & \longrightarrow \ \ L^2 ( {\br}, dp)
\\
 & f & \longmapsto \ \  {\cal F} f
\quad {\rm with} \ \;
\left( {\cal F} f \right) (p)
= {1 \over \sqrt{2\pi \hbar}} \int_{\br} dx \, f(x) \,
{\rm exp} (-{\ri \over \hbar}px )
\ \ .
\nonumber
\end{eqnarray}

\subsection{Operators}

In this section, we will properly define the notion of an 
operator and its adjoint (the Hermitian conjugate operator), 
as well as Hermitian operators 
and self-adjoint operators (i.e. physical 
observables).
Furthermore, we will indicate how the properties of the spectrum 
of a given operator are intimately connected 
with the mathematical characteristics of this operator and 
of its eigenfunctions.   
 We strongly encourage the 
reader who is not familiar with the 
definitions and results spelled out 
at the beginning of this subsection 
to pursue the reading 
with the illustrations that follow.

The following considerations (i.e. sections 2.2.2 and 2.2.3) 
do not only apply to 
the concrete Hilbert space $\lt$ of wave mechanics, but to any  
Hilbert space $\hs$ that is isomorphic to it,
i.e. any complex Hilbert space which admits a countably
infinite basis\footnote{The precise definition of an 
isomorphism is recalled in section 3.2 below.}; 
therefore, we will refer to a general Hilbert space
$\hs$ of this type and only specialize to the 
concrete realizations $\lt$
or $L^2 ([0,1], dx)$ in the examples.

Just as a function $f : \br \to \br$ 
has a domain of definition 
${\cal D}(f) \subset \br$,  
a Hilbert space operator $A : \hs \to \hs$ 
admits such a domain: 
\begin{defin}
An {\em operator on the Hilbert space $\hs$} is a linear map 
\begin{eqnarray}
A \ : & {\cal D} (A) & \longrightarrow \ \ \hs
\\
 & \psi  & \longmapsto \ \
A  \psi 
\ \ ,
\nonumber
\end{eqnarray}
 where ${\cal D}(A)$ represents a dense linear subspace of $\hs$.
This subspace is called 
 the {\em  domain of definition of} $A$
or, for short, the {\em  domain of} $A$.  
(Thus, strictly speaking, a 
Hilbert space operator
is a pair $(A, {\cal D}(A)$) consisting of 
a prescription of operation on the Hilbert space, together with 
a Hilbert space subset on which this operation is 
defined.)
 
If $B$ denotes another operator on $\hs$ 
$($with domain ${\cal D} (B))$, then the operator $A$ 
is said to be {\em equal} 
to the operator $B$ if both 
the prescription of operation and the domain
of definition coincide, i.e. if
\[
A \varphi =  B \varphi 
\qquad \mbox{for all} \ \varphi \in {\cal D} (A) = {\cal D} (B)
\ \ . 
\]
In this case, one writes $A=B$. 
\end{defin}
We note that, in the mathematical literature,
the definition of an operator 
is usually phrased in more general terms 
by dropping the 
assumptions that ${\cal D} (A)$ is dense and that the map $A$
is linear; however, these generalizations scarcely occur 
in quantum mechanics and therefore we will not consider them here. 

As emphasized in the previous definition, 
an operator is not simply a formal operating 
prescription and two operators
which act in the same way are to be considered 
as different if they
are not defined on the same subspace of Hilbert space.
A typical example for the latter situation is given by  
a physical problem on a compact or semi-infinite interval: the 
domain of definition of operators then includes, in general,
some boundary conditions whose choice 
depends on the experimental set-up which 
is being considered. 
Clearly, two non equivalent experimental set-ups 
for the measurement of a given physical observable 
generally lead to different experimental results:
thus, it is important to consider as different
two Hilbert space operators which act in the same way,
but admit different domains of definition.
Actually, most of the contradictions derived 
in section  2 can be traced back to the 
fact that the domains of definition
of the operators
under consideration have been ignored!

By way of example, let us first consider 
three different operators on the Hilbert space $\hs =\lt$ 
and then another one on $L^2([0,1])$. 

\medskip 
\noindent
{\bf 1.a)}
The {\em position operator} $Q$ for a particle 
on the real line 
is the operator 
`multiplication by $x$'
on $\lt$:
\begin{equation}
\label{pos}
\left(Q \psi \right) (x) = x \, \psi (x)
\quad  \mbox{for all $x\in \br$}
\ \ .
\end{equation}
The {\em maximal domain of definition} for $Q$ is 
the one which ensures that the function 
$Q\psi$ exists and that it still belongs to the Hilbert
space : 
\begin{eqnarray}
{\cal D}_{\rm max} (Q) 
&=& \{ \psi \in \hs \ | \ Q\psi \in \hs \}
\nonumber 
\\
\label{dq}
&=& \{ \psi \in \lt \ | \
\| Q \psi \|^2 \equiv \int_{\br}dx \, x^2 | \psi (x) |^2
< \infty \}
\ \ .
\end{eqnarray}
Obviously, this represents a proper subspace of $\lt$ and it can 
be shown that it is a dense subspace \cite{sg,rs}.  

\medskip 
\noindent
{\bf 1.b)}
Analogously, 
the maximal domain of definition of the {\em momentum operator} 
$P= \ds{\hbar \over \ri} \, \ds{d \over dx}$ 
on the Hilbert space $\lt$
is\footnote{Since the integral involved in the definition of the space 
$\lt$ is the one of 
Lebesgue, one only needs to ensure that the considered functions 
behave correctly `almost everywhere' with respect to Lebesgue's
measure (see textbooks on analysis): 
thus, $\psi^{\prime} \in \lt$ means that the derivative 
$\psi^{\prime}$ exists almost everywhere and that it belongs to 
$\lt$.}
\[
{\cal D}_{\rm max} (P) = \{ \psi \in \lt \, | \,
\psi^{\prime} \in \lt \}
\ \ .
\]

\medskip 
\noindent
{\bf 1.c)}
For certain considerations it is convenient to 
have at one's disposal 
a domain of definition that is left invariant by the operator
(rather than a maximally defined operator).   
For the operator $Q$, such a domain is given by the Schwartz space 
$\st$ of rapidly decreasing functions. 
Let us recall that a function $f : \br \to \bc$ belongs to $\st$
if it is differentiable an infinite number of times and if this function,  
as well as all of its derivatives, 
decreases more rapidly at infinity than  
the inverse of any polynomial). This implies that 
$\st \subset {\cal D}_{\rm max} (Q)$ and 
\begin{equation}
\label{qsc}
Q \ : \ \st \longrightarrow \st
\ \ .
\end{equation}

The Schwartz space also represents an 
{\em invariant domain of definition}
for the momentum operator 
$P = \ds{\hbar \over \ri} \, \ds{d \over
dx}$ on $\lt$, i.e. 
$P \, : \, \st \rightarrow \st$.

According to the definition given above, the operators $P$ and $Q$
introduced in the last example are  different
from $P$ and $Q$ as defined in the 
preceding two examples, respectively; however, 
in the present case where we have an infinite interval,  
this difference is rather 
 a mathematical one since
physical measurements do not allow us to make  
a distinction between the different domains. 

\medskip 
\noindent
{\bf 2.} 
Let us also give an example concerning a particle 
which is confined  
to a compact interval e.g. $[0,1 ]$. 
In this case, the wave 
function generally satisfies some boundary conditions
which have to be taken into account by means of the 
domains of the operators under investigation. 
For instance, let us consider example 4 of section 2, 
i.e. the momentum operator $P$ on  
$\hs =  L^2 ([0,1], dx)$ 
with the usual boundary conditions 
of an infinite potential well: 
the domain of definition of $P$ then reads  
\begin{equation}
\label{dir}
{\cal D} (P) = \{ \psi \in \hs \, | \,
\psi^{\prime} \in \hs \ {\rm and} \ \psi(0) = 0 = \psi(1) \}
\ \ .
\end{equation} 
Shortly, we will come to the mathematical and physical 
properties of this quantum mechanical operator. 

\bigskip 

The specification of the domain 
plays a crucial role when  
introducing the adjoint $A^{\dag}$ of a
Hilbert space operator
$A$:

\begin{defin}
For an operator $A$ on $\hs$, the 
{\em domain of} $A^{\dag}$
is defined by 
\begin{eqnarray*}
{\cal D} (A^{\dag}) & = & \{ \varphi \in \hs \, | \, \exists\,
\tilde{\varphi} \in \hs \ \ \mbox{such that}
\\
&& \qquad \qquad \qquad \qquad \quad
\left\langle \varphi , A \psi \right\rangle =
\langle \, \tilde \varphi  \, , \, \psi \, \rangle
\ \mbox{for all} \ \psi  \in {\cal D} (A) \}
\ \ .
\end{eqnarray*}
(Here, the vector $\tilde{\varphi}$
depends on both $A$ and  $\varphi$.)
For $\varphi \in {\cal D}(A^{\dag})$, one defines
$A^{\dag} \varphi = \tilde \varphi$, i.e.
\begin{equation}
\label{defa}
\langle \varphi , A \psi \rangle =
\langle A^{\dag} \varphi , \psi \rangle
\qquad \mbox{for all} \ \psi \in {\cal D}(A)
\ \ .
\end{equation}
\end{defin}

By way of example, we again consider the Hilbert space 
$\hs =  L^2 ([0,1], dx)$
and the momentum operator 
$P$ with the domain (\ref{dir}). According to the previous definition, 
the domain of $P^{\dag}$ is given by 
\[
{\cal D} (P^{\dag}) = \{ \varphi \in \hs \, | \, \exists\,
\tilde{\varphi} \in \hs \ \mbox{such that} \
\langle \varphi , P \psi \rangle =
\langle \tilde \varphi , \psi \rangle
\ \mbox{for all} \ \psi  \in {\cal D} (P) \}
\]
and the operating prescription for $P^{\dag}$ is determined  
by the relation 
\begin{equation}
\langle \varphi , P \psi \rangle =
\langle P^{\dag} \varphi , \psi \rangle
\qquad \mbox{for all} \ \psi \in {\cal D}(P)
\ \ .
\end{equation}
The integration by parts (\ref{itg})
or, more precisely, 
\begin{eqnarray}
\int_0^1 \! d x \, (  \overline{\varphi} \, P\psi
-
\left( \overline{\ds{\hbar \over \ri} \,
\ds{d\varphi \over dx} }  \right) \psi ) ( x )
&=& \ds{\hbar \over \ri} \left[ \overline{\varphi( 1)} \psi (1) -
\overline{ \varphi ( 0)} \psi (0) \right] 
\nonumber 
\\
& =& 0
\qquad  \mbox{for all} \ \, \psi \in {\cal D}( P)
\nonumber 
\end{eqnarray} 
shows that the boundary conditions satisfied by 
$\psi \in {\cal D}(P)$ are already sufficient 
for annihilating the surface term
and it shows that $P^{\dag}$ acts in the same way as  $P$. 
Hence, 
\begin{equation}
\label{dadj}
P^{\dag}  = \ds{\hbar \over \ri} \, \ds{d \over dx}
\quad , \quad
{\cal D} (P^{\dag}) = \{ \varphi \in \hs \, | \, \varphi^{\prime}
\in \hs \}
\ \ .
\end{equation}
Thus, the domain of definition of $P^{\dag}$ is larger 
than the one of $P$:
${\cal D}(P) \subset {\cal D}(P^{\dag})$.
 
\bigskip 

In quantum theory,  
physical observables are described by Hilbert space operators 
which have the property of being self-adjoint.
Although many physics textbooks use 
this term as a synonym for Hermitian,
there exists a subtle difference between these two 
properties for operators acting on {\em in}finite dimensional 
Hilbert spaces: as we will illustrate in the sequel,
this difference 
is important for quantum physics.  
\begin{defin}
The operator $A$ on $\hs$ is {\em Hermitian} if 
\begin{equation}
\langle \varphi , A \psi \rangle  =
\langle A \varphi , \psi \rangle
\qquad \mbox{for all} \ \varphi, \psi \in {\cal D}(A)
\end{equation}
i.e.  
\[
A^{\dag} \varphi = A  \varphi
\qquad \mbox{for all} \ \varphi \in {\cal D}(A)
\ \ .
\]
(In other words, the operator $A$ is Hermitian if $A^{\dag}$
acts in the same way as $A$ on all vectors belonging to ${\cal D}(A)$,
though $A^{\dag}$ may actually be defined on a larger 
subspace  than ${\cal D}(A)$.) 
 
An operator $A$ on $\hs$ is
{\em self-adjoint} if the operators 
$A$ and $A^{\dag}$  coincide ($A= A^{\dag}$), i.e. 
explicitly  
\begin{equation}
{\cal D} (A) = {\cal D} (A^{\dag})
\qquad 
and \qquad 
A^{\dag} \varphi = A  \varphi
\quad \mbox{for all} \ \varphi \in {\cal D}(A)
\ \ .
\end{equation}
\end{defin}
Thus, any self-adjoint operator is Hermitian, but 
an Hermitian operator is not necessarily self-adjoint.
Our previous example provides an  
illustration for the latter fact. 
We found that $P$ and $P^{\dag}$ act in the same way, though 
the domain of definition 
of $P^{\dag}$ (as given by equation (\ref{dadj}))
is strictly larger than the one of $P$ 
(as given by equation (\ref{dir})):
${\cal D}(P) \subset {\cal D}(P^{\dag})$, but 
${\cal D}(P) \neq {\cal D}(P^{\dag})$.
From these facts, we conclude that $P$ is Hermitian, 
but not self-adjoint:
$P \neq P^{\dag}$. 

One may wonder whether it is possible to characterize in another 
way the little ``extra'' that a 
Hermitian operator is lacking in order to  
be self-adjoint. This missing item 
is exhibited by the following
result which is proven in mathematical textbooks. 
If the Hilbert space operator $A$ is self-adjoint, then  
its spectrum is real \cite{ri,sg}\cite{af}-\cite{krey} and 
the eigenvectors associated to different eigenvalues 
are mutually orthogonal; moreover, the eigenvectors together 
with the generalized eigenvectors
 yield a {\em complete} 
system of (generalized) vectors of the Hilbert space\footnote{The 
vectors which satisfy the eigenvalue equation, though they 
do not belong to the Hilbert space $\hs$, but to a larger space 
containing $\hs$, are usually referred to as 
``generalized eigenvectors'', 
see section 2.2.3 below.} 
\cite{gv,bere,sg}.
These results do not hold for operators which are 
only Hermitian. 
As we saw in section 2.1, 
this fact is confirmed by our previous example:
the Hermitian operator $P$ does not admit any 
proper or generalized eigenfunctions 
and therefore it is not self-adjoint (as we already 
deduced by referring directly to the 
 definition of self-adjointness).  

Concerning these aspects, we recall 
that the existence of a complete 
 system of (generalized)
eigenfunctions is fundamental 
 for the physical interpretation of 
observables. It motivates the mathematical definition of an observable 
which is usually given in textbooks on 
quantum mechanics: an  
{\em observable} is defined as a ``Hermitian operator 
whose orthonormalized eigenvectors define a basis of Hilbert 
space" \cite{ct}. 
The shortcomings of this approach (as opposed to 
the identification of observables with self-adjoint operators)
will be discussed in section 4. Here, we only note 
the following. 
For a given Hilbert space operator, it is usually 
 easy to check whether it is Hermitian (e.g. by performing
 some integration by parts). 
And for a Hermitian operator, 
there exist 
simple criteria for self-adjointness: the 
standard one will be 
stated and applied in the appendix. 

Finally, we come to the spectrum of operators
for which we limit ourselves to the general ideas. 
By definition, the {\em spectrum of a self-adjoint operator} 
$A$ on the Hilbert space $\hs$ 
is the union of two sets of real numbers, 
\begin{enumerate}
\item
the so-called {\em discrete} or {\em point spectrum},
i.e. the set of eigenvalues of $A$
(that is eigenvalues for which 
the eigenvectors belong to the domain of definition 
of $A$).
\item
the so-called {\em continuous spectrum},
i.e. the set of generalized eigenvalues of $A$ (that is 
eigenvalues for which ``the eigenvectors do not belong to 
the Hilbert space $\hs$'').
\end{enumerate}
These notions will be made more precise 
and illustrated in the next subsection as well as in the appendix. 
From the physical point of view, the spectral values of an  
observable (given by a self-adjoint operator)  
are the possible results
that one can find upon measuring this physical quantity. 
The following two examples are familiar from physics.

\medskip 

\noindent
{\bf 1.}
For a particle moving on a line, the observables of position and
momentum can both take any real value. Thus, the corresponding 
operators have a purely continuous and unbounded spectrum:
${\rm Sp} \, Q = \br$ and ${\rm Sp} \, P = \br$. 
A rigorous proof of this result will be given shortly
(equations (\ref{vpq})-(\ref{dis}) below).

\medskip 

\noindent
{\bf 2.}
For a particle confined to the unit interval and subject to 
periodic boundary conditions, we have 
wave functions belonging to the Hilbert space 
$L^2 ( [0,1],dx)$ 
and satisfying the boundary conditions 
$\psi (0) = \psi (1)$. 
In this case, the position operator 
admits a continuous and bounded spectrum given by the interval  
$[0,1]$. The spectrum of the momentum operator 
is discrete and unbounded, 
which means that the momentum can only take certain discrete, though 
arbitrary large values. 

\medskip 

In order to avoid misunderstandings, we should emphasize that the
definitions given in physics and mathematics 
textbooks, respectively, for the spectrum of an operator acting on 
an infinite dimensional Hilbert space, do not completely coincide.
In fact, from a mathematical point of view it is not  
natural to simply define 
the {\em spectrum of a generic Hilbert space operator} as 
the set of its proper and generalized eigenvalues (e.g. see 
references \cite{jau} and \cite{sg}):
by definition, it contains a third part, the so-called 
{\em residual spectrum}. The latter 
is empty for self-adjoint operators  
and therefore we did not mention it above. 
However, we will see in the appendix that 
the residual spectrum is not empty 
for operators which are only Hermitian; 
we will illustrate its usefulness for 
deciding whether a given 
Hermitian operator can possibly be modified so as to become 
self-adjoint
(i.e. precisely the criteria for self-adjointness 
of a Hermitian operator that we  
already evoked). 

Before concluding our overview of Hilbert space operators, 
we note that all definitions
given in this subsection 
also hold if the complex Hilbert space $\hs$
is of finite dimension $n$, i.e. $\hs \simeq \bc^n$.
In this case (which is of physical interest for  
the spin of a particle or for a quantum mechanical two-level system),  
the previous results simplify greatly: 

- any operator $A$ on $\hs$  
(which may now be expressed in terms of a complex $n\times n$ matrix)
and its adjoint (the Hermitian conjugate matrix) 
are  defined on the whole Hilbert space, 
i.e. ${\cal D}(A) = \hs$

- Hermitian and self-adjoint are now synonymous

- the spectrum 
of $A$ is simply the set of all its eigenvalues, i.e. we have  
 a purely discrete spectrum.    

Yet, it is plausible that the passage to infinite dimensions, i.e. 
the addition of 
an infinite number of orthogonal directions 
to $\bc^n$ 
opens up new possibilities!
As S.MacLane put it: ``Gentlemen: there's lots of room 
left in Hilbert space.'' \cite{rs}

\subsection{Observables and generalized eigenfunctions}

When introducing the observables of position and momentum
for a particle on the real line, we already noticed that they 
are not defined on the entire Hilbert space, but 
only on a proper subspace of it. 
In this section, we will point out and discuss 
some generic features
of quantum mechanical observables.

Two technical complications appear in the study 
of an observable $A$
in quantum mechanics:
 
\noindent (i) If the spectrum of $A$ is not bounded, then 
the domain of definition of $A$ cannot be all of $\hs$.
 
\noindent (ii) If the spectrum of $A$ contains a continuous part,  
then the corresponding eigenvectors 
do not belong to $\hs$, but rather to a larger space.  
 
Let us discuss 
these two problems in turn before concluding with some related remarks.  

\subsubsection{(i) Unbounded operators}

The simplest class of operators is the one of {\em bounded} 
operators, i.e. for every vector 
$\psi \in {\cal D}(A)$, one has 
\begin{equation}
\label{bor}
\| A  \psi \| \leq c \, \| \psi \|
\qquad \mbox{where $c\geq 0$ is a constant}
\ \ .
\end{equation}
This condition amounts to say that the spectrum of $A$ is bounded. 
Bounded operators can always be defined on the entire Hilbert space, 
i.e. ${\cal D}(A)= \hs$. 
 An important example is the one of an unitary operator 
$U : \hs \to \hs$ ; such an operator is bounded, because
it is norm-preserving 
(i.e. $\| U  \psi \| = \| \psi \|$
 for all $\psi \in \hs$) and therefore condition (\ref{bor})
is satisfied. (The spectrum of $U$ lies
on the unit circle of the complex plane and therefore it is bounded.
For instance, 
the spectrum of the Fourier transform operator ${\cal F}$ 
defined by (\ref{f}) is the discrete set $\{ \pm 1, \pm \ri \}$ 
\cite{af,lio}.)
 
A large part of the mathematical subtleties of quantum mechanics
originates from the following result
\cite{sg, rs}.
\begin{theo}[Hellinger-Toeplitz]
   Let  $A$ be an operator on $\hs$ which is everywhere defined 
   and which satisfies the Hermiticity condition 
\begin{equation}
\label{her}
\langle \varphi , A  \psi \rangle =
\langle A  \varphi  ,   \psi \rangle
\end{equation}
for all vectors $\varphi, \psi \in \hs$.
Then $A$ is bounded.
\end{theo}
In quantum theory, one often deals with operators, like those 
associated to the position, momentum or energy, which fulfill the 
Hermiticity condition (\ref{her}) on their domain of definition,
but for which the spectrum is not bounded. 
(In fact, the basic structural relation of quantum mechanics, 
i.e. the canonical commutation relation, even imposes 
that some of the fundamental operators, which are involved in it,   
are unbounded - see appendix.) 
The preceding theorem
indicates that it is not possible to define these Hermitian 
operators on the entire Hilbert space $\hs$ and that their
domain of definition {\em necessarily} 
represents a proper subspace of $\hs$.
Among all the choices of subspace which are possible from 
the mathematical point of view, certain ones are usually privileged 
by physical considerations (boundary conditions, ...) 
\cite{aw,rs,sg,ber,th}.

By way of example, we consider the position operator 
$Q$ defined on the Schwartz space $\st$, see 
eqs.(\ref{pos})(\ref{qsc}). 
This operator is Hermitian since all vectors 
$\varphi, \psi \in \st$ satisfy
\[
\langle \varphi , Q \psi \rangle 
= \int_{\br} dx \, \overline{\varphi} \, x \psi 
= \int_{\br} dx \, \overline{x \varphi} \, \psi 
= \langle Q \varphi ,  \psi \rangle
\ \ . 
\]
As we are going to prove shortly, the spectrum of 
this operator is the entire real axis (reflecting the fact that 
$Q$ is not bounded). 
For this specific operator, we already noticed  
explicitly that there is no way to define $Q$  
on all vectors of Hilbert space: at best, it can be defined 
on its maximal domain (\ref{dq})
which represents a nontrivial subspace of Hilbert space.

\subsubsection{(ii) Generalized eigenfunctions (Gelfand triplets)}
 
The position operator Q defined on $\st$ also illustrates
the fact that the eigenvectors associated to the continuous spectrum  
of a self-adjoint operator do not belong to the Hilbert 
space\footnote{To be precise, the operator $Q$ 
defined on $\st$ is {\it essentially self-adjoint}
which implies that it can be rendered self-adjoint in a unique manner 
by enlarging its domain of definition in a natural way 
(see \cite{rs, sg} for details).}.
In fact, the eigenfunction $\psi_{x_0}$ associated to the 
eigenvalue $x_0 \in \br$ is defined by the 
relation
\begin{equation}
\label{vpq}
\left( Q \psi_{x_0} \right) (x) = x_0 \, \psi_{x_0} (x)
\qquad (x_0 \in \br \ ,  \ \psi_{x_0} \in \st \ , \
\psi_{x_0} \not \equiv 0 )
\end{equation}
or else, following (\ref{pos}), by 
\[
(x-x_0) \, \psi_{x_0} (x) = 0
\quad  \mbox{for all  $x\in \br$}
\ \ .
\]
 This condition implies 
$\psi_{x_0} (x) =0$ for $x \neq x_0$. 
Consequently, the function $\psi_{x_0}$ vanishes almost 
everywhere\footnote{Actually, $\psi_{x_0} \in \st$
implies that $\psi_{x_0}$ is continuous and therefore 
it follows that $\psi_{x_0}$ vanishes everywhere.} 
and thus represents the null vector of 
$\lt$ \cite{sg, af, rs}. 
 Hence, the operator $Q$ does not admit any eigenvalue:
its discrete spectrum is empty. 

We note that the situation is the same for the operator $P$
defined on $\st$ which is also essentially self-adjoint: 
the eigenvalue equation 
\[
\left( P\psi_p \right) (x) = p \, \psi_p (x)
\qquad ( p \in \br \ , \ \psi_p \in \st \ , \
\psi_p \not \equiv 0 ),
\ \ ,
\]
is solved by $\psi_p (x) = 1/\sqrt{2\pi \hbar}
\  {\rm exp} (\ri px / \hbar )$, but 
$\psi_p \not \in \st$.
Thus $P$ does not admit any eigenvalue. 

On the other hand, the eigenvalue equations for $Q$ and $P$
admit weak (distributional) solutions. 
For instance, Dirac's generalized function (distribution) 
with support in $x_0$, i.e. 
$\delta_{x_0} (x) \equiv \delta (x-x_0)$,
is a weak solution of the eigenvalue equation (\ref{vpq}):
in order to check that 
$x \, \delta _{x_0} (x) = x_0 \, \delta_{x_0} (x)$
in the sense of 
distributions, we have to smear out this relation with a test function 
$\varphi \in \st$:
\begin{equation}
\label{for}
\int_{\br} dx \,
x \, \delta _{x_0} (x) \, \varphi (x) = x_0 \, \varphi (x_0) =
\int_{\br} dx \,
x_0 \, \delta _{x_0} (x) \, \varphi (x)
\ \ .
\end{equation}
Dirac's generalized function and the generalized function
$x \delta _{x_0}$ do not belong to the domain of definition 
$\st$ of $Q$, rather they belong to its dual space 
\[
\std = \{ \omega : \st \to {\bf C} \
\mbox{linear and continuous} \}
\ \ ,
\]
i.e. the space of tempered distributions on 
$\br$ \cite{rs, sg, bgc, ls, gv}.
They are defined in an abstract and rigorous manner by  
\begin{eqnarray}
\label{xd}
\delta_{x_0} \ : & \st  & \longrightarrow \ \ \bc
\\
 & \varphi & \longmapsto \ \
\delta_{x_0} (\varphi ) = \varphi (x_0)
\nonumber
\end{eqnarray}
and
\begin{eqnarray*}
x\, \delta_{x_0} \ : & \st  & \longrightarrow \ \ \bc
\\
 & \varphi & \longmapsto \ \
\left( x\, \delta_{x_0} \right) (\varphi ) = \delta_{x_0} (x \varphi )
\stackrel{(\ref{xd})}{=}
x_0 \, \varphi (x_0)
\ \ .
\end{eqnarray*}
With these definitions, the formal writing (\ref{for})
takes the precise form 
\[
\left(
x \, \delta _{x_0} \right) (\varphi) = \left(
x_0 \, \delta _{x_0} \right) (\varphi)
\qquad \mbox{for all} \ \ \varphi \in \st
\ \ .
\]
Thus, the eigenvalue equation $Q \psi_{x_0} = x_0 \, \psi_{x_0}$
admits a distributional solution $\psi_{x_0}$ for every value 
$x_0 \in \br$.
Since the spectrum of the (essentially self-adjoint) operator 
$Q$ is the set of all real numbers for which the eigenvalue equation admits
as solution either a function $\psi \in {\cal D}(Q) = \st$ 
{\em (discrete spectrum)} 
or a generalized function $\psi \in \std$ 
{\em (continuous spectrum)}, 
we can conclude that ${\rm Sp} \, Q = \br$ and that 
the spectrum of $Q$ is purely continuous.
 
Analogously, the function 
$\psi_p (x) = 1/\sqrt{2\pi \hbar}
\  {\rm exp} ( \ri px / \hbar)$
defines a distribution $l_p$ according to 
\begin{eqnarray}
\label{dis}
l_p  \ : & \st  & \longrightarrow \ \ \bc
\\
 & \varphi & \longmapsto \ \
l_p (\varphi ) = \int_{\br} dx \, \overline{\psi_p (x)} \,
\varphi (x) 
\stackrel{(\ref{f})}{=} 
({\cal F} \varphi ) (p)
\ \ ,
\nonumber
\end{eqnarray}
 where ${\cal F} \varphi$ 
 denotes the Fourier transform 
(\ref{f}).
The distribution $l_p$ represents a solution of the eigenvalue equation 
$Pl_p = p \, l_p$ since the 
calculational rules for distributions and Fourier transforms
\cite{sg,gv} and the definition (\ref{dis}) imply that 
\[
\left( Pl_p \right) (\varphi) =
\left( \ds{\hbar \over \ri} \, \ds{dl_p \over dx} \right) (\varphi)
=
l_p \left( \ds{\hbar \over \ri} \, \ds{d\varphi \over dx} \right) 
\stackrel{(\ref{dis})}{=}
( {\cal F}
\left( \ds{\hbar \over \ri} \, \ds{d\varphi \over dx} \right) ) (p)
= p \, ( {\cal F} \varphi ) (p) 
\stackrel{(\ref{dis})}{=}
p \, l_p (\varphi)
\ \ .
\]
It follows that 
${\rm Sp} \, P = \br$ (purely continuous spectrum).
 
The eigenvalue problem for the operators $P$ and $Q$ which admit a
continuous spectrum 
thus leads us to consider the following 
{\em Gelfand triplet}
({\em ``rigged} or  {\em equipped 
Hilbert space"})\footnote{Gelfand triplets 
are discussed in detail in the textbook \cite{gv}
(see also \cite{bere} and \cite{jr} 
for a slightly modified definition).
A short and excellent introduction to the definitions 
and applications in quantum mechanics is given in references 
\cite{ber,blt,sg}. Concerning the importance
of Gelfand triplets, we cite their 
inventors \cite{gv}: ``We believe that this concept is no less 
(if indeed not more) 
important than that of a Hilbert space."}
\begin{equation}
\label{tri}
\fbox{\mbox{$ \
\st \subset \lt \subset \std \ $}}
\ \ .
\end{equation}
Here, $\st$ is a dense subspace of $\lt$ \cite{rs}
and every function 
$\psi \in \lt$ defines a distribution $\omega_{\psi} \in \std$
according to 
\begin{eqnarray}
\omega_{\psi} \ : & \st & \longrightarrow \ \ {\bf C}
\nonumber
\\
 & \varphi & \longmapsto \ \
\omega_{\psi} (\varphi) = \int_{\br} dx \, \overline{\psi(x)}
\varphi (x)
\ \ .
\label{reg}
\end{eqnarray}
 However $\std$ also contains distributions  
 like Dirac's 
  distribution $\delta_{x_0}$ 
or the distribution $l_p$ which cannot be represented 
by means of a function $\psi \in \lt$ according to (\ref{reg}).
The procedure of smearing out with a test function 
$\varphi \in \st$ corresponds 
to the formation of {\em wave packets} and the theory of 
distributions gives a quite precise meaning to this procedure 
as well as to the generalized functions that it involves.   

The abstract definition of the triplet (\ref{tri}) 
can be made more precise (e.g. specification of the 
topology on $\st$,...) and,  
furthermore, $\st$ can be generalized to other 
subspaces of $\lt$ 
(associated to $Q$ or to other operators defined on 
$\lt$). 
{\bf Up to these details} (which are important and 
which have to be taken 
into account in the study of 
a given Hilbert space operator), 
we can say that
\begin{quote}
{\bf the triplet (\ref{tri})
describes in an exact and simple manner the mathematical nature 
of all kets and bras used 
in quantum mechanics.}\footnote{The reader who is not acquainted 
with bras and kets may skip this paragraph 
and come back to it after browsing through section 3.}  
\end{quote}
   In fact, according to the Riesz lemma (which is 
recalled in section 3.1.1 below), 
the Hilbert space $\lt$ is isometric  
to its dual: thus, to each ket 
belonging to $\lt$, there corresponds a bra 
given by an element of $\lt$,
and conversely. 
Moreover, a ket belonging to the subspace $\st$ always defines a bra
belonging to $\std$ by virtue of definition (\ref{reg}).
But there exist elements of $\std$, the {\em generalized bras}, 
to which one cannot associate a ket belonging to $\st$ or $\lt$.
We note that the transparency of this mathematical result 
is lost if one proceeds as one 
usually does in quantum mechanics, namely if one 
describes the action 
of a distribution on a test function $\varphi$ in a purely formal 
manner as a scalar product between 
$\varphi \in \st \subset \lt$
and a function which does not belong to $\lt$:
\begin{eqnarray}
\nonumber 
l_p (\varphi ) 
\! & \! = \! & \! 
\left\langle \psi_p ,\varphi \right\rangle _{L^2}
\equiv \int_{\br} dx \, \overline{\psi_p (x)} \, \varphi (x)
\qquad (\, \mbox{ignoring that}\ \psi_p \not \in \lt \,)
\\
\label{ign}
\delta_{x_0} (\varphi )
\! & \! = \! & \! 
\left\langle \delta_{x_0} , \varphi \right\rangle _{L^2}
\equiv \int_{\br} dx \, \overline{\delta_{x_0} (x)} \, \varphi (x)
\quad
\ \ (\, \mbox{ignoring that}\ \delta_{x_0} \not \in \lt \,).
\end{eqnarray}
 
Quite generally, let us consider a self-adjoint operator 
$A$ on the Hilbert space $\hs$. The eigenfunctions 
associated to elements of the continuous spectrum of $A$
do not belong to the Hilbert space $\hs$: one has to equip 
$\hs$ with an appropriate dense subspace $\Omega$ and its dual 
$\Omega^{\prime}$ which contains the generalized eigenvectors
of $A$, 
\[
\Omega \subset \hs \subset \Omega^{\prime}
\ \ .
\]
The choice of the subspace $\Omega$ is intimately connected 
 with the domain of definition of the operator $A$
one wants to study\footnote{Generally speaking, one has to choose
the space $\Omega$ in a maximal way so as  
to ensure that $\Omega^{\prime}$ is as ``near'' as possible
to $\hs$ : this provides the closest possible 
analogy with the finite-dimensional case where 
$\Omega = \hs = \Omega^{\prime}$
\cite{bere,ber}. In fact, if one chooses $\Omega$ large enough,
the space $\Omega^{\prime}$ is so small that 
the generalized eigenvectors of $A$ which belong to 
$\Omega^{\prime}$ 
``properly'' characterize 
this operator (see references \cite{ber,bere} for
a detailed discussion).}. 
While the introduction of the space  $\Omega$
is mandatory for having a well-posed mathematical problem, 
the one of $\Omega^{\prime}$ is quite convenient, though not 
indispensable for the determination of the spectrum of $A$.
In fact, there exist several {\em characterizations of the spectrum} 
which do not call for an extension of the Hilbert space\footnote{In 
this context, let us cite the authors of reference \cite{rs}: 
``We only recommend the abstract
rigged space approach to readers with a strong emotional attachment
to the Dirac formalism." This somewhat provocative statement 
reflects fairly well the approach followed in the majority of
textbooks on functional analysis.}.
Let us mention three examples.
The different parts of the spectrum of $A$
can be described by different properties of the {\em resolvent}
$R_A (z) = (A - z {\bf 1})^{-1}$
(where $z \in \bc$) \cite{sg,bgc} or  (in the case where $A$
is self-adjoint) by the properties of the {\em spectral projectors}
$E_A (\lambda)$ (where $\lambda \in \br$)
associated to $A$
\cite{jvn, sg, rs} 
or else by replacing the notion of distributional eigenfunction 
of $A$ by the one of {\em approximate eigenfunction}
\cite{bgc}. (The latter approach reflects the well-known fact that 
distributions like $\delta_{x_0}$ can be approximated arbitrarily well 
by ordinary, continuous functions.)

\subsubsection{(iii) Concluding remarks} 

As we already indicated, the examples discussed in the previous subsection 
exhibit both the problems raised by an unbounded 
spectrum and by 
a continuous part of the spectrum. 
We should 
emphasize that these problems are not related to each other;
this is illustrated by the position and momentum operators for 
a particle confined to a compact interval and described by a wave function
 with periodic boundary conditions 
(see example 2 at the end of section 2.2.2).

We note that it is sometimes convenient to pass over from unbounded
to bounded operators. One way to do so consists of using exponentiation
to pass from self-adjoint to unitary operators (which are necessarily 
bounded). 
For instance, if $Q$ and $P$ denote the (essentially) self-adjoint
operators of position and momentum 
with the common domain of definition 
$\st$, then 
$U_a \equiv {\rm exp} \, ( -{\rmi \over \hbar} a Q)$ and 
$V_b \equiv {\rm exp} \, ( -{\rmi \over \hbar} b P)$ (with $a,b \in \br$)
 define one-parameter families of unitary operators. 
In particular, $V_b$ represents the translation operator for 
wave functions:
\[
\left( V_b \psi \right) (x) = \psi (x-b) 
\qquad {\rm for} \ \, \psi \in \lt
\ \ .
\]
This bounded operator
admits a continuous spectrum consisting of the unit circle.  
If written in terms of $U_a$ and $V_b$, 
the canonical commutation relation (CCR)
$[P,Q ] = \ds{\hbar \over \rmi} {\bf 1}$ takes the so-called
{\em Weyl form} 
\begin{equation}
U_a V_b = \rme ^{-{\rmi \over \hbar} ab} V_b U_a
\ \ .
\end{equation}   
This relation can be discussed
without worrying about domains of definition
since it only involves bounded operators.
It was used by J. von Neumann to prove his famous 
{\em uniquess theorem for the representations of the CCR:} 
this result states that the 
Schr\"odinger representation 
$`Q= \mbox{multiplication by} \ x', 
\;  P = \ds{\hbar \over \ri}\ds{d \over dx}$ 
is essentially the only possible realization of the CCR \cite{rs,gap}.

\chapter{Quantum mechanics and Hilbert spaces}
 
 The different representations used in quantum mechanics are
 discussed in numerous monographs \cite{ct} and we will 
 summarize them in the next subsection 
while evoking the historical  development. 
 The underlying mathematical theory is expounded in textbooks
 on functional analysis  \cite{af}. Among these books, there are some
 excellent monographs which present the general theory together with its
 applications to quantum mechanics
 \cite{rs,sg,bgc} (see also \cite{ri,lio,tj,krey}).
The basic result which is of interest to us here
(and on which we will elaborate in subsection 3.2)
is the fact that all representations mentioned above 
are equivalent. 
This implies that the problems we presented 
in section 2 are not simply artefacts of wave mechanics
and that such problems manifest themselves 
 in quantum theory  
whatever formulation is chosen to describe the theory. 
In fact, we could have written some of the 
equations  explicitly in terms of operators 
$P, \, Q, \, L_z, \, \varphi , ...$
acting  on an abstract Hilbert space and 
in terms of vectors 
$| m \rangle ,...$ belonging to such a space, e.g. eqs.(\ref{hei})
and  (\ref{lzp})-(\ref{inc}).

   \section{The different Hilbert spaces}

As in section 2.2, we consider the motion of a particle 
on a straight line parametrized by 
$x \in {\bf R}$. 
There are essentially three Hilbert spaces which
are used for the description of the states  of this 
particle.
 
\bigskip
 
\noindent
{\bf (1) Wave mechanics:} 

\medskip 

Wave mechanics was developed in 1926 by E.~Schr\"odinger
``during a late erotic outburst in his life''  \cite{pais,moore}. 
There were two sources of 
inspiration. The first was the concept of matter waves which 
L.~de Broglie had introduced in his thesis 
a few years earlier (1923). The second was given by the papers
 on statistical gas theory that 
Einstein had published in 1925 with the aim of extending 
some earlier work 
of S.N.~Bose (which papers led to the famous 
``Bose-Einstein statistics'' for indistinguishable 
particles):
Einstein believed that molecules as well as photons must 
have both particle and wave characters \cite{moore}.
E.~Schr\"odinger laid the foundations of wave mechanics 
in three remarkable papers
that he wrote 
in early 1926, just after returning from 
the studious Christmas holidays 
that he had spent with an unknown companion in  
the ski resort of Arosa \cite{moore}.    
Looked upon in retrospective and 
from the mathematical point of view, 
he considered the Hilbert space $\lt$
of square integrable functions (i.e. wave
functions) with the scalar product (\ref{ip}) 
(or, equivalently, the space 
$L^2 (\br, dp)$ of wave functions depending on the momentum
$p$, i.e. the so-called {\em $p$-representation}). 
It is worthwhile recalling that 
any element $\psi \in \lt$ can be expanded with respect 
to a given orthonormal basis 
 $\{ \varphi_n \} _{n\in \bn}$ of $\lt$:
\begin{equation}
\label{exp}
\psi = \sum_{n =0}^{\infty} c_n  \varphi_n  
\qquad {\rm with} \ \;
c_n = \langle \varphi_n , \psi \rangle _{L^2} \in \bc 
\ \ .
\end{equation}
If $\psi^{\prime} = \sum_{n} c_n^{\prime}  \varphi_n$ 
denotes another element of  $\lt$, its scalar product
with $\psi$ takes the form 
\begin{equation}  
\label{exun}
\langle \psi , \psi^{\prime} \rangle _{L^2}
= \sum _{n  =0}^{\infty} \bar c_n c_n ^{\prime}
\ \ .
\end{equation}
In particular, for the norm $\| \psi \| 
\equiv \sqrt{\langle \psi , \psi \rangle}$ of $\psi$, 
we have 
\begin{equation}  
\label{norma} 
\| \psi \| ^2 = 
\sum _{n  =0}^{\infty} | c_n | ^2 
\ \ .
\end{equation}

\noindent {\bf (2) Matrix mechanics:}

\medskip 

Once an orthonormal basis  
$\{ \varphi_n \} _{n\in \bn}$ of $\lt$ has been chosen, 
the expansion (\ref{exp}) of $\psi \in \lt$ 
with respect to this basis  
expresses a one-to-one correspondence between the 
wave function $\psi$ and the infinite sequence  
$\vec c = (c_0, c_1,...)$ of complex numbers.
By virtue of relation (\ref{norma}), 
the {\em square integrability} of the function $\psi$ 
(i.e. $\| \psi \|^2 \equiv \int_{\br} | \psi (x)|^2 dx 
< \infty$) 
is equivalent 
to the {\em square summability} of the sequence $\vec c$
(i.e. $\sum_n | c_n |^2 < \infty$).  

Thus, quantum mechanics can also  be formulated 
using the Hilbert space $l_2$ 
of square summable sequences,  
$$
l_2 = \{ \vec{c} = (c_0, c_1, ...) \mid c_n \in {\bf C}
\ \ {\rm and} \ \; \sum_{n=0}^{\infty} \ | c_n | ^2 < \infty \}
\ \ ,
$$
with the scalar product 
$$
\langle \vec c , \vec c ^{\; \prime} \, \rangle _{l_2} 
= \sum_{n=0} ^{\infty}
\overline{c_n} \, c^{\prime} _n
\qquad \mbox{for all} \ \; 
\vec{c}, \vec c ^{\; \prime} \in l_2 
\ \ .
$$
The linear operators on $l_2$ are then given 
by square matrices of infinite order.
This formulation of quantum theory is known as matrix mechanics
and was actually the first one to be discovered 
\cite{pais, kragh, moore}:

- In 1925, Heisenberg wrote his pioneering paper pointing 
in the right direction. 

- A first comprehensive account
of the foundations of matrix mechanics 
was given in November 1925 in the famous ``Dreim\"annerarbeit"
(three men's work) of Born, Heisenberg and Jordan. 

- Some of these results have been put forward    
independently by Dirac 
who introduced the 
concept of 
{\em commutator} of operators 
and identified it as the quantum analogon 
of the classical Poisson brackets \cite{wig}. 
 
- Pauli's subsequent derivation of the 
spectrum of the hydrogen atom using matrix methods  
provided strong evidence for the correctness of this approach.  

Despite this success, matrix mechanics
was still lacking clearly stated physical 
principles and therefore only represented high
mathematical technology. 
The situation changed in 1926 after Schr\"odinger
proved the equivalence between wave mechanics and 
matrix mechanics, which is 
based on the correspondence
\begin{equation}
\label{corr}
\left.
\begin{array}{r}
\mbox{wave function}
\\
 \psi \in \lt 
\end{array}
\right\}
 \ \ \stackrel{{\rm 1-1}}{\longleftrightarrow}
\ \ 
\left\{
\begin{array}{l}
\mbox{sequence}
\\
\vec c \in l_2  
\ \ .
\end{array}
\right.
\end{equation}
This result 
led Born to propose his probabilistic interpretation 
of wave functions 
which finally provided the physical principles 
of the theory. 
The equivalence (\ref{corr}) also 
represented the starting point for the search of an ``invariant"
version of quantum mechanics. Through the work of Dirac and Jordan,
this undertaking led to the study of linear operators acting on
an abstract Hilbert space \cite{fh}\cite{jvn,stone}.

It is a remarkable coincidence that the monograph of Courant and
Hilbert  \cite{ch} developing the mathematics of  Hilbert space was
published in 1924 and that it appeared to be written 
specifically for the physicists of this time\footnote{D.~Hilbert:
``I developed my theory of
infinitely many variables from purely mathematical interests
and even called it `spectral analysis' without any
pressentiment that it would later find an application
to the actual spectrum of physics." \cite{cr}}.
Incidentally, 
the space $l_2$ has been introduced  
in 1912 by D.Hilbert 
in his work on integral equations, but an axiomatic 
definition of Hilbert space was only given in 1927 by 
J.~von Neumann in a paper on the mathematical foundation 
of quantum mechanics \cite{krey}. 
In the sequel, the theory of Hilbert space operators 
was further  refined, mainly through  the
contributions of 
von Neumann \cite{jvn}, Schwartz \cite{ls} and 
Gelfand \cite{gv}. Thanks to these refinements, it allows 
for a precise description of all states and observables in quantum
mechanics. 
To conclude our historical excursion on the mathematical 
formulation of quantum theory, we recall the remarkable fact 
that some of its main protagonists
(namely Heisenberg, Dirac,
Jordan, Pauli  and von Neumann) achieved their  
ground-breaking contributions to physics and mathematics 
around the age of 23.

\bigskip
 
\noindent {\bf (3)
The invariant formalism:} [Dirac, Jordan, von Neumann,
1926 - 1931]

\medskip 

One uses an abstract complex Hilbert space $\hs$
 that is {\em separable}
(which means that it admits an orthonormal basis consisting 
of a denumerable family of vectors)
and {\em infinite dimensional}.  

It was in this framework that Paul Adrien Maurice 
Dirac invented the astute
bra and ket notation, following a particular
interpretation of the expressions involving vectors and 
operators\footnote{As pointed out by H.Kragh \cite{kragh}, 
a peculiar feature 
of Dirac's physics was his interest 
in notation and his readiness to invent new terms 
and symbols. 
Apart from the bra and ket notation, the 
$\delta$-function and the commutator of operators 
that he introduced, we should mention that he coined 
the terms $c$-number, fermion and boson 
and that he defined $\hbar$ as a short form 
for $h /2 \pi$.}.    
Before discussing 
this interpretation and the resulting advantages and shortcomings 
(section 4.2), we first  
summarize Dirac's notation together
with some fundamental concepts  
of the theory of Hilbert spaces \cite{ri,sg}\cite{af}-\cite{krey}.

\subsection{Dirac's notation}

A vector $|\Psi \rangle \in \hs$ is called a {\em ket}$\,$
and to this vector we can associate a linear form 
$\omega_{|\Psi \rangle} \equiv \langle \Psi |$
called a {\em bra}$\,$ and defined by means of the scalar product
($\!${\em ``bracket"}$\,$):
\begin{eqnarray}
\label{bra}
\omega_{|\Psi \rangle} \equiv \langle \Psi |
\ : & \hs & \stackrel{{\rm lin.}}{\longrightarrow} \ \ {\bf C}
\\
 & |\Phi \rangle & \longmapsto \ \
\omega_{|\Psi \rangle} \left( | \Phi \rangle \right)
= \langle \,
|\Psi \rangle \, , \,   | \Phi \rangle \, \rangle  _{\hs} \equiv
\langle \Psi | \Phi \rangle
\ \ .
\nonumber
\end{eqnarray}
According to the Cauchy-Schwarz inequality relating 
the scalar product 
and the norm 
$\|  \Psi  \| \equiv \| |  \Psi  \rangle \| =
\sqrt{\langle \Psi | \Psi \rangle}$
in $\hs$,
\[
| \langle \Psi | \Phi \rangle | \leq
\| \Psi \|   \cdot
\| \Phi  \|
\ \ ,
\]
the linear form 
$\omega_{|\Psi \rangle}$ is {\em continuous} on $\hs$: 
this means that for every 
$|\Phi \rangle \in \hs$ 
there exists a constant 
$c\geq 0$ 
such that 
$| \omega_{| \Psi \rangle} (| \Phi \rangle ) | \leq
c  \, \| \Phi \|$. 
Consequently, the bra 
$\langle \Psi |$ is an element of the {\em dual Hilbert space} 
\[
\hs^{\ast} = \{ \omega : \hs
\longrightarrow {\bf C} \ \; \mbox{linear and continuous} \}
\ \ .
\]
 
Conversely, to each bra $\langle \Psi | \in \hs^{\ast}$
we can associate a ket $| \Psi \rangle \in \hs$;
in fact, by virtue of the {\em Riesz lemma},
every element  $\omega \in \hs^{\ast}$ 
uniquely determines a vector $| \Psi _{\omega} \rangle
\in \hs$ such that 
\[
\omega \left( | \Phi \rangle \right)
= \langle \Psi_{\omega} | \Phi \rangle
\qquad \mbox{for all} \quad | \Phi \rangle \in \hs
\ \ .
\]
(The vector 
$| \Psi _{\omega} \rangle$ ``realizes" the map $\omega$
by means of the scalar product.)  
The vector associated to the linear form $\langle \Psi |$
is denoted by $| \Psi \rangle$ and thus we have a 
one-to-one correspondence between $| \Psi \rangle \in \hs$
and $\langle \Psi | \in \hs^{\ast}$:
\begin{equation}
\fbox{\mbox{$ \ \hs \ni
| \Psi \rangle \ \ \stackrel{{\rm 1-1}}{\longleftrightarrow}
\ \ \langle \Psi | \in \hs^{\ast} \ $}}
\ \ .
\end{equation}
Hence we can identify\footnote{If one defines the norm of  
$\omega \in \hs^{\ast}$ by $\| \omega \| =
{\rm sup} \, | \omega (f) |$ (where the supremum is taken 
 over all unit vectors $f \in \hs$), then one can show that the 
bijection $\hs \to \hs^{\ast}$ is antilinear and that it is 
norm-preserving, i.e. it represents an isometry.} $\hs$ and $\hs^{\ast}$    
and completely do without $\hs^{\ast}$. 
The introduction of a dual vector space is only necessary for defining 
generalized vectors, as we saw in section 2.2.3. 

An {\em orthonormal basis} 
$\left\{ | \Phi_n \rangle \equiv |n\rangle \right\}_{n\in \bn}$
 of $\hs$ is a set of vectors satisfying the orthonormalization 
 relation 
 \begin{equation}
\langle n | m \rangle = \delta_{nm}
\qquad \mbox{for all} \quad n,m \in \bn
\end{equation}
and the closure relation 
\begin{equation}
\label{id}
\sum_{n=0}^{\infty} | n \rangle \langle n | =  {\bf 1}_{\hs}
\ \ .
\end{equation}
Applying the latter to any vector $| \Psi \rangle \in \hs$, 
we find the expansion 
\[
| \Psi \rangle = \sum_{n=0}^{\infty} | n \rangle 
\langle n | \Psi \rangle
\ \ ,
\]
that we encountered 
already for wave functions, cf. eq.(\ref{exp}).
Note that the relation (\ref{id}) 
involves the sum of the operators 
$| n \rangle \langle n |$ 
which are obtained by composing two maps: 
\begin{eqnarray}
\label{de}
\hs \ \ \stackrel{\langle n|}{\longrightarrow} & {\bf C}&
\stackrel{|n \rangle}{\longrightarrow} \quad \ \ \hs
\\
| \Phi \rangle \ \longmapsto & \langle n | \Phi \rangle &
\longmapsto \
\langle n | \Phi \rangle \,  |n \rangle
\ \ .
\nonumber
\end{eqnarray}
Here, the first map is the linear form 
(\ref{bra}) and the second represents the multiplication of a complex
number by the vector $|n\rangle \in \hs$.
 
As stated in section 2.2.2, 
an operator on $\hs$ is a linear map 
\begin{eqnarray}
A \ : & {\cal D} (A) & \longrightarrow \ \ \hs
\\
 & |\Psi \rangle & \longmapsto \ \
A | \Psi \rangle
\ \ ,
\nonumber
\end{eqnarray}
 where ${\cal D}(A)$ 
is a dense linear subspace of $\hs$.
The scalar product of the vectors 
$|\Phi \rangle$ and $A | \Psi
\rangle$ is denoted following Dirac by 
\begin{equation}
\label{mat}
\langle \, | \Phi \rangle \, ,\, A|\Psi \rangle \, \rangle _{\hs} \equiv
\langle \Phi | A | \Psi \rangle
\ \ .
\end{equation}
Thus, the expression $\langle \Phi | A | \Psi \rangle$ 
may be considered as the result of the 
composition of two linear maps, 
\begin{eqnarray}
\label{copo}
{\cal D} (A) \ \ \stackrel{A}{\longrightarrow} & \hs &
\stackrel{\langle \Phi |}{\longrightarrow} \quad \ \ {\bf C}
\\
| \Psi \rangle \ \longmapsto & A | \Psi \rangle &
\longmapsto \
\langle \Phi | A | \Psi \rangle
\ \ ,
\nonumber
\end{eqnarray}
where the composition is defined as usual by 
$\left( \langle \Phi | \circ A \right)  | \Psi \rangle :=
\langle \Phi | \left(  A   | \Psi \rangle \right)$.
However, as we will discuss further in section 4.2.1,
Dirac did not restrict himself to 
this unambiguous 
interpretation of the notation 
that he introduced.

\section{Relations between the Hilbert spaces}
 
 In order to describe the relations between the Hilbert spaces
 introduced in the previous section, we need 
 the concepts of unitary operator and of isomorphism \cite{rs}.
By way of motivation, we recall 
that equation (\ref{corr}) expresses a one-to-one correspondence 
between the Hilbert spaces of wave mechanics and of matrix mechanics, 
and that equation (\ref{exun}) implies that 
this correspondence preserves scalar products,
i.e. it is realized by a unitary operator:
this is an example of an 
isomorphism of Hilbert spaces.

\begin{defin}
For $i=1,2$, let $\hs_i$  be a complex separable Hilbert space with
scalar product  $\langle \ , \ \rangle _{\hs_i}$. 
A linear operator
$U : \hs_1 \to \hs_2$ is called  {\em unitary}
if
 
\noindent {\rm (i)} $U$ is everywhere defined on $\hs_1$.
 
\noindent {\rm (ii)} The image of $\hs_1$ under $U$ is all of 
$\hs_2$.
 
\noindent {\rm (iii)} $U$ preserves the scalar product :
\begin{equation}
\label{un}
\langle U f , U g \rangle _{\hs_2} = \langle f , g \rangle _{\hs_1}
\qquad \mbox{for  all} \quad f,g \in \hs_1
\ \ .
\end{equation}
 
Two Hilbert spaces $\hs_1$ and  $\hs_2$ which are related by 
 an unitary operator are said to be {\em isomorphic} and one writes 
$\hs_1 \simeq \hs_2$.
\end{defin}
	
To summarize, we can say that two isomorphic Hilbert spaces 
represent different realizations of the same abstract 
structure and that they can be considered as completely equivalent. 

Concerning the Hilbert spaces occurring in quantum mechanics,
we have at our disposal a  classic result of functional analysis : 

\begin{theo}
{\rm (i)} The complex Hilbert spaces $l_2, \,  \lt$ and  $\ltp$
are separable and infinite dimensional. 
 
\noindent
{\rm (ii)} Every complex Hilbert space which is separable and infinite
dimensional is isomorphic to $l_2$.
\end{theo}

From this result it follows that all Hilbert spaces
introduced above are isomorphic : 
\begin{equation}
\fbox{\mbox{$
\hs \simeq l_2 \simeq \lt \simeq \ltp$}}
\ \ .
\end{equation}
In particular, 
the {\em Parseval-Plancherel theorem} states that the 
Fourier transformation (\ref{f}) realizes the 
isomorphism between $\lt$ and $\ltp$ :
\[
\langle {\cal F} f, {\cal F} g \rangle = 
\langle  f,  g \rangle
\qquad \mbox{for all} \ \; f,g \in \lt
\ \ .
\]

Quite generally, the passage between $\hs$
and the other Hilbert 
spaces is performed by choosing an orthonormal basis 
$\{ | n \rangle \}_{n \in \bn}$ (or a generalized 
basis $\{ | x \rangle \}_{x \in \br}$)
of $\hs$ and by associating to each vector of $| \Psi \rangle \in \hs$
the set of its components with respect to this basis: 
\begin{eqnarray}
\psi_n & \! \! :=  \! \! & \langle n | \Psi \rangle 
\quad {\rm for} \ \; n \in \bn 
\qquad , \quad \vec \psi \equiv (\psi_0 , \psi_1, ...)  \in l_2 
\nonumber 
\\
\label{xr}
\psi (x) & \! \! := \! \! & \langle x | \Psi \rangle 
\quad {\rm for} \ \; x \in \br 
\qquad \, , \quad \psi \in \lt 
\ \ .
\end{eqnarray}
In the second line, the action of $\langle x |$ on  
$| \Psi \rangle$ is to be understood in the sense of the action 
of a distribution on a test vector 
- see equations (\ref{xd})(\ref{ign}). 
(Thus, strictly speaking, $| \Psi \rangle$ has to belong to 
an appropriate  test vector subspace $\Omega \subset \hs$.) 
Schematically:

\bigskip 

\begin{picture}(500,100)(0,0)
\put(263,90){\makebox(0,0){$| \Psi \rangle \in \hs$}}
\put(332,10){\makebox(0,0){$\vec \psi \in l_2 $}}
\put(195,10){\makebox(0,0){$\psi \in L^2$}}
\put(260,75){\vector(1,-1){50}}
\put(240,75){\vector(-1,-1){50}}
\put(200,60){\makebox(0,0){$\langle x |$}}
\put(299,60){\makebox(0,0){$\langle n |$}}
\end{picture}

\bigskip

The passage between  $\lt$ and $l_2$ is realized in an analogous
 manner and was already described above:
 to the function $\psi \in \lt$, one associates 
the sequence 
   $(c_0 , c_1, ...)  \in l_2 $ consisting of the components 
      $c_n :=  \langle \varphi_n , \psi \rangle_{L^2}$ 
of $\psi$ 
with respect to an orthonormal basis $\{ \varphi_n \}_{n\in \bn}$
of $\lt$.
If the wave functions $\varphi_n$
are associated to the 
state vectors $| n \rangle \in \hs$, i.e.  
$\varphi_n (x) =
 \langle x | n \rangle$, 
then consistency 
with the notation (\ref{xr})
requires that $c_n = \psi_n$, i.e. 
$\langle \varphi_n , \psi \rangle_{L^2} =
\langle n | \Psi \rangle$.

\chapter{Discussion of the invariant formalism 
and of Dirac's notation}

In this section, we try to disentangle two matters
which are generally presented in an interwoven way, 
namely the choice of an abstract Hilbert space 
for formulating quantum mechanics
and 
the use of the bra and ket notation.
By discussing the advantages and disadvantages
of each of these choices, we aim to 
highlight their positive points, which one should try 
to preserve in practice, and 
to draw attention to the pitfalls 
which are to be avoided.

\section{The invariant formalism}

Since the different Hilbert spaces used in quantum mechanics
are all isomorphic, they represent  
completely equivalent mathematical structures. 
However, from the practical point of view, certain spaces are 
more appropriate than others\footnote{G.Orwell:
``All animals are equal, but some animals are more equal
than others."}.
\begin{enumerate}
\item
The matrix calculus based on the space $l_2$
is not easy to handle and this formalism has barely been used after 
the advent of quantum mechanics for which it played an 
important role \cite{vdw}.
\item
The arena of physical phenomena is the so-called 
configuration space parametrized by $x$, and boundary or regularity
conditions directly concern the 
wave functions defined on this space: 
this privileges the use of the Hilbert space $\lt$. 
\item
The choice of an abstract Hilbert space $\hs$ is usually motivated by  
the analogy with geometry in Euclidean space 
$\br^n$ (or ${\bf C}^n$): the use of ``abstract vectors" is more geometrical
than the one of their components. Thus, it is tempting to work with the 
vectors $ | \Psi \rangle \in \hs$
while interpreting the sequences belonging to $l_2$ or the functions 
belonging to $\lt$ as the components of the vectors 
$|\Psi \rangle $ with respect to different bases of $\hs$.
In this spirit, the use of an abstract Hilbert space in quantum mechanics 
is often presented as something  more general than wave or matrix mechanics 
\cite{ct}.
However, there are crucial differences between finite and
infinite dimensional vector spaces which render the analogy with 
ordinary geometry quite subtle and doubtful.  
In the following, we will discuss the resulting problems 
which show that the choice of an abstract Hilbert space in 
quantum mechanics obscures or complicates 
important points of the theory. 
\end{enumerate}
 
\subsection{Problems}
 
\begin{itemize}
\item
For the study of simple problems like the determination of 
the energy spectrum of the harmonic oscillator 
(for which the eigenfunctions of the Hamiltonian are 
well defined elements of $\lt$),
one has to start with the introduction of the 
eigendistributions $|x \rangle$ of the position operator 
which do not belong to the Hilbert space 
$\hs$ (section 2.2.3).
\item
As we emphasized in section 2.2.2 and illustrated in the appendix, 
the definition of a linear operator on an infinite dimensional 
Hilbert space necessitates the specification of an 
operating prescription \underline{and} of a domain of definition 
for this operation. 
This aspect does not simply represent a mathematical subtlety, 
since the spectrum of the operator is quite sensitive to the 
domain of definition (boundary conditions,...). 
For instance, depending on the choice of domain,
the spectrum of the momentum operator 
$\ds{\hbar \over \ri} \ds{d \over dx}$ on a compact interval
$[a,b] \subset \br$ can be 
empty, all of ${\bf C}$ or a subset of 
 $\br$ (see appendix  and reference \cite{sg}).
While this problem is well posed,  from the beginning on, for wave
functions defined on configuration space, it is not 
to the same extent for an abstract Hilbert space.

This problematics also appears in quantum statistical mechanics, 
e.g. the definition of the pressure associated to a 
 set of particles confined to a box involves the 
boundary conditions \cite{rob}.
\item 
Let us come back to the mathematical characterization 
of observables which was already touched upon in section 2.2.2. 

In the invariant formalism of quantum mechanics, an 
{\em observable} is defined as an ``Hermitian operator 
whose orthonormalized eigenvectors define a basis of Hilbert 
space" \cite{ct}. 
Starting from this definition, it can then be shown, in a formal 
manner, that the position and momentum operators on $\br$ are 
observables. 
Notable complications already occur in $\br^2$ or $\br^3$
if non-Cartesian coordinates 
are considered, 
e.g. the radial component 
$P_r$ of momentum in $\br^3$ is Hermitian, but it does not represent 
an observable - see Messiah \cite{ct} chap.9. 
And there are distinctly more complicated operators 
like Hamiltonians involving random potentials, potentials
of the form $1/  x^n$ or $\delta_{x_0} (x)$
or else Hamiltonians on topologically nontrivial configuration 
spaces like those for the Aharonov-Bohm effect or for anyons
\cite{rs, bs, ber}.
The definition of an observable given above then requires 
to impose {\em ad hoc} conditions on the wave functions 
associated to eigenstates (conditions of regularity, finiteness,
single-valuedness,...); furthermore, it necessitates the
explicit determination  of an orthonormal system of eigenvectors and the
verification  of the closure relation for this system. 

In an approach which takes into account the domains of definition, 
an observable is simply given by a self-adjoint operator 
(section 2.2.2). This condition ensures that 
the spectrum of the operator is real and that its (generalized)
eigenvectors define a (generalized) basis of Hilbert space
{\em (``Hilbert's spectral theorem")}.
Moreover, there exist simple criteria for checking
whether a given Hermitian 
operator is self-adjoint, or for classifying 
the different manners according to 
which it can be rendered self-adjoint
- see \cite{rs,sg,th} and appendix. (In general, 
if an operator admits several self-adjoint extensions, 
the latter describe different physical situations
\cite{aw,rs}.) In particular, it is not necessary 
to resort to some {\em ad hoc} properties of wave functions like those 
mentioned above or to try to determine a complete system of 
orthonormal eigenvectors. 
The relevance of such a simple and precise approach 
also comes to light in perturbation \cite{th,klau}
or scattering theory \cite{amj}. 
\item
An important concept of quantum mechanics is the one of CSCO
(complete system of commuting observables). 
It involves the commutativity of self-adjoint operators 
which represents a subtle notion for unbounded operators. 
In fact, two self-adjoint operators $A$ and $B$ commute if and only if 
all projection operators occurring in their respective 
spectral decompositions commute \cite{rs}.
Unfortunately, counterexamples show that, in order to have  
the commutativity of $A$ and $B$, 
it is not enough that $[A,B] =0$ on a
dense  subspace of $\hs$ on which this relation 
is well defined \cite{rs}!
Admittedly, these examples ever scarcely  appear in practice, 
but, in an approach which takes into account the domains
of definition, one has at ones disposal all the tools 
that have to be called for, if a mathematical complication 
manifests itself. 
\end{itemize}
Concerning the raised mathematical points, we emphasize that 
in quantum mechanics a precise formulation is not  
only required for deciding about the existence or non-existence
of physical effects (like the Aharonov-Bohm effect \cite{ab}),
but also for discussing the difficult interpretational problems
(measurement theory, objectivity and reality,...) \cite{au}. 
Besides, such a formulation 
directly applies to other fields of physics, one example being 
chaos in classical dynamical systems \cite{pri}.
 
\subsection{``Solution" of problems}

Certain of the problems pointed out in the previous subsection 
are so involved that it seems more advisable to avoid them, rather 
than to 
look for a remedy. The complications mainly arise from the fact that 
- for conceptual reasons - one wishes to put forward 
the geometric Hilbert space structure which is underlying the theory.
But this structure is also implicit in wave mechanics
where the raised problems are {\em absent} or, at least,  
{\em well posed from the beginning on}. 
Thus, it is easy to avoid mathematical troubles or at least 
to render them more transparent by working with wave mechanics. 

However, one cannot ignore that 
the abstract Hilbert space formalism is 
so widely used in the physics literature 
that all physicists need to be acquainted  
with it, though, at the same time, they should 
be able to resolve mathematical puzzles like 
those  presented in section 2; 
thus,  for the teaching of quantum mechanics, an obvious 
``solution" of problems is to start with an introduction 
to wave mechanics which emphasizes the underlying geometric 
structures and to indicate the arbitrariness of this formulation 
by passing over to other representations like matrix mechanics.
(Such an approach to quantum theory and Schr\"odinger operators 
is supported by a vast and easily accessible 
mathematical literature.)
In modifying the explicit definition of the Hilbert space and its 
scalar product, the formalism of wave mechanics on  $\br$
can then be generalized straightforwardly
to several spatial dimensions, 
to the spin or to systems of particles.
A subsequent discussion of the abstract Hilbert space $\hs$ 
and its relation to $\lt$ then paves the way to the literature
and to formal calculations.

\section{Dirac's notation}
 
As we mentioned already, Dirac's bra and ket formalism 
consists, on one hand, of a certain writing of vectors, linear forms, ...
and, on the other hand, of a particular interpretation of the 
mathematical operations which involve these entities.

\subsection{Inconveniences}

This writing, or rather its interpretation, 
presents a certain number of drawbacks 
which are more or less embarrassing. 
Among these, the most troublesome is the fact that it is 
{\em impossible} to give a precise meaning to the adjoint 
$A^{\dag}$ (of an unbounded operator $A$) if one strictly adheres
to Dirac's interpretation (see \cite{grau} and also \cite{fano}).
Concerning this aspect, let us recall Dirac's fundamental definition 
(e.g. see equations (B.45)
and (B.51) of chapter II of Cohen-Tannoudji et al. \cite{ct}):
\begin{equation}
\label{cohe}
\left( \langle \Phi | A \right) | \Psi \rangle =
\langle \Phi | \left( A  | \Psi \rangle \right) \equiv
\langle \Phi |  A  | \Psi \rangle
\qquad {\rm with} \quad
\langle \Phi | A  = \langle A^{\dag} \Phi |
\ \ .
\end{equation}
According to these relations, one cannot tell whether 
the expression 
$\langle \Phi |  A  | \Psi \rangle$ 
is to be interpreted as
\[
\langle  A^{\dag} \Phi | \Psi \rangle
\qquad
(\, \mbox{in which case} \ \; |\Phi \rangle \in {\cal D} (A^{\dag})
\ \; {\rm and} \ \; | \Psi \rangle \in \hs \, )
\]
or as 
\[
\langle \Phi | A   \Psi \rangle
\qquad
(\, \mbox{in which case} \ \; |\Psi \rangle \in {\cal D} (A) \
\; {\rm and} \ \; | \Phi \rangle \in \hs \, ),
\]
unless one reintroduces the parentheses (which obviously 
takes away the simplicity and elegance of the calculus). 
Alas \cite{rs}, it is possible that 
${\cal D}(A^{\dag}) = \{ 0 \}$
for an operator $A$ which is defined on a dense subspace of $\hs$.
Even though this case scarcely ever appears in practice, 
the examples 3
and 7 of section 2 show that the ignorance of
domains of definition can 
readily lead to contradictions and incorrect results; 
accordingly, the 
correct treatment of a problem involving operators 
which cannot be defined everywhere (unbounded operators)
is subtle. 

If one agrees upon the assumption that  
$\langle \Phi |  A  | \Psi \rangle$ 
 is to be interpreted as 
$\langle \Phi | \left( A  | \Psi \rangle \right)$, so that 
$\langle \Phi |A$ does not stand for $\langle A^{\dag} \Phi |$, 
but simply for a composition of operators
(according to equation (\ref{copo})), 
then the mathematical ambiguities concerning matrix elements are 
discarded. Yet, some 
inconveniences remain: 
we will discuss these in the 
familiar case where one {\em strictly} 
applies the bra and ket notation 
in an abstract, infinite dimensional Hilbert space $\hs$. 
 
\begin{itemize}
 
\item {\em Rigid notation:}
Let us first recall the standard definition of the 
adjoint $A^{\dag}$ of a linear operator\footnote{In order to avoid 
the discussion of domains of definition, we assume that $A$ is 
a bounded operator: in this case, $A$ and $A^{\dag}$ can  
be defined on the entire Hilbert space $\hs$.}
$A : \hs \to \hs$ :
\begin{equation}
\label{ad}
\langle \, | \Phi \rangle \, , \, A | \Psi \rangle \, \rangle _{\hs}
\, = \,
\langle \, A^{\dag} | \Phi \rangle \, ,\, |\Psi \rangle \, \rangle _{\hs}
\qquad 
\mbox{for all} \ \; 
|\Phi \rangle , |\Psi \rangle \in \hs 
\ \ .
\end{equation}
 If one rigidly adheres to Dirac's notation, the 
expression on the right-hand side has to be 
rewritten using the skew-symmetry of the scalar product,
$\langle \Phi | \Psi \rangle =
\langle \Psi | \Phi \rangle ^{\ast}$; thus, relation
(\ref{ad}) which defines the adjoint of $A$ becomes 
\begin{equation}
\langle \Phi | A | \Psi \rangle  =
\langle \Psi | A^{\dag} | \Phi \rangle ^{\ast}
\qquad \mbox{for all} \ \; 
|\Phi \rangle , |\Psi \rangle \in \hs 
\ \ .
\label{adj}
\end{equation}
Consequently, the matrix element 
$\langle A | \Phi \rangle , | \Psi \rangle \, \rangle _{\hs}$ 
can only be represented by 
$\langle \Psi | A | \Phi \rangle ^{\ast}$ or by 
$\langle \Phi | A^{\dag} | \Psi \rangle$.
A frequently used example is given by 
\[
\| A | \Psi \rangle \|^2 \; = \;
\langle A | \Psi \rangle , A | \Psi \rangle \rangle _{\hs} \; =\;
\langle \, |\Psi \rangle ,A^{\dag} A|\Psi \rangle \rangle _{\hs}
 \; \equiv \; \langle \Psi |A^{\dag} A |\Psi \rangle
\ \ ,
\]
where the last expression is the only acceptable writing according
to Dirac. 
 
\item
{\em Lack of naturalness and simplicity:}
As indicated in section 3.1.1, 
one can do without the discussion of the 
dual space $\hs^{\ast}$ (the space of bras), 
since this one is isometric to $\hs$. 
Now Dirac's formalism makes a systematic use of $\hs^{\ast}$.
While we are accustomed to operators and matrices acting 
on ``everything in front of them", in this formalism 
one has to distinguish between the action of linear operators 
to the right and to the left
\cite{ct},
\begin{eqnarray}
A \left( \lambda | \Phi \rangle +  \mu | \Psi \rangle \right)
& = &
\lambda A | \Phi \rangle +  \mu A | \Psi \rangle
\qquad (\lambda , \mu \in {\bf C} )
\nonumber
\\
\label{aad}
\left( \lambda \langle \Phi | + \mu \langle \Psi | \right) A
& = &
\lambda \langle \Phi | A +
\mu \langle \Psi | A
\ \ ,
\end{eqnarray}
which entails potential ambiguities concerning the
domains of definition. 
Furthermore, one has to change the natural order of vectors 
in certain expressions that are often used 
(e.g. compare equations  (\ref{ad}) and (\ref{adj})). 
 
\item
{\em Changing computational rules:}
When passing from $\hs$ to $\lt$
(which one is practically always obliged to do at a certain point, 
since physics is happening in configuration space), 
some of the calculational rules change: the operators 
of differentiation on $\lt$ only act to the right, 
their matrix elements can be written as 
$\langle A \varphi , \psi \rangle _{L^2}$, ... and so on. 
 
\item {\em Difficult mathematical interpretation in an abstract 
Hilbert space $\hs$:}
If one assumes (as we did) that the vectors belong to 
an abstract, infinite dimensional Hilbert space, 
then one recovers all of the problems mentioned 
in section 4.1. 
In this case, Dirac's bra and ket formalism represents a 
{\bf purely symbolic calculus} and it is certainly not 
by chance that von Neumann did not look for an explanation 
or mathematical formulation of this approach 
when working out the mathematical foundations of quantum mechanics
\cite{jvn} (see also \cite{stone}). 
The modern introductions to this formalism 
try to render its mathematical content a bit more precise, 
but there exist only a few serious attempts which try to 
translate Dirac's approach into a rigorous mathematical theory, 
following an appropriate interpretation of it 
\cite{jr,jpa,ein,gl,jau}.
The resulting theories (most of which involve,  
from the beginning on,  
abstract Gelfand triplets and spectral families) 
are quite complicated and difficult to handle. 
 
\item
{\em Non pedagogical approach:}
The basic concepts of linear algebra (like linear maps, 
scalar products,...) are heavily used in all fields of physics 
(analytical mechanics, electrodynamics, relativity,...) 
with the standard mathematical notation and not with Dirac's 
formalism. Functional analysis based on the space 
$\lt$ (or on $l_2$) is a natural synthesis of linear algebra 
and real analysis and some notions of this theory are part
of the standard mathematical luggage of any physicist
(e.g. by means of Fourier analysis). 
On the other hand, Dirac's symbolic calculus sometimes conveys  
the impression of representing something qualitatively new 
and essentially unavoidable for the development of quantum 
mechanics\footnote{It may be worthwhile to recall that quantum 
theory has been developed without the use of this formalism 
\cite{vdw} and to note that its teaching can largely or 
completely do without it, as demonstrated by a fair number of excellent 
textbooks \cite{gap,ll,pd}.}.
\end{itemize}

For the finite dimensional Hilbert spaces 
involved in the description of the spin (or the angular momentum) of
particles, one has $\hs \simeq {\bf C}^n$ and Dirac's notation 
then represents a rewriting of vectors and 
a particular interpretation of the operations 
of standard linear algebra \cite{fano}.
In this case, all expressions are mathematically well defined, 
but not all of the other inconveniences that we mentioned are 
discarded.

\subsection{Advantages}
 
 The great power of Dirac's notation consists of the fact that 
it allows us to perform formal calculations 
  which automatically lead to the correct form of the results. 
  For instance, insertion of the identity operator (\ref{id})
  between two operators $A$ and $B$,
\begin{equation}
\label{in1}
\langle \Phi | AB| \Psi \rangle 
\stackrel{(\ref{id})}{=}
\langle \Phi | A
\left( \sum_{n=0}^{\infty} |n \rangle
\langle n | \right)  B| \Psi \rangle
= \sum_{n=0}^{\infty}
\langle \Phi | A| n \rangle
\langle n | B| \Psi \rangle
\ \ ,
\end{equation}
immediately yields the right form of the 
final result without the need
to contemplate the successive action of the maps 
$|n \rangle$ and $\langle n |$ described in equation 
(\ref{de}).
Similarly, the projection $P_{|\Phi \rangle}$
on the state $|\Phi \rangle \in \hs$ simply reads
\begin{equation}
\label{in2}
P_{|\Phi \rangle} =
|\Phi \rangle \langle \Phi |
\]
and for its matrix elements, one readily gets  
\[
\langle \Psi | P_{|\Phi \rangle} | \Psi^{\prime} \rangle =
\langle \Psi | \Phi \rangle  \langle \Phi | \Psi^{\prime} \rangle
\ \ .
\end{equation}

\subsection{``Solution" of problems}
 
As we just emphasized, the notation $| \Psi \rangle$
for vectors and $\langle \Psi |$ for linear forms is 
quite useful for mnemonic and 
computational purposes. 
Thus, it would be out of place to avoid this notation and to 
do without its advantages. 
A good compromise which we will now summarize is the one adopted 
or mentioned in a certain number of texts \cite{gap,ll,pd,th}.
 
Since the concrete Hilbert space $\lt$ is isomorphic to the 
abstract Hilbert space $\hs$
of Dirac's approach, we can 
``identify" them: by writing  
$\hs = \lt$, we already avoid the mathematical 
complications of the invariant formalism (section~4.1). 
In any case - whether or not this identification is made - 
it is often convenient to use Dirac's bra and ket notation 
within wave mechanics. Henceforth, ones 
{\em writes the wave functions as $| \psi \rangle$}
rather than $\psi$
- or as $\, \psi \rangle$ as suggested by Dirac 
himself \cite{d} 
- in order to memorize the following relations which hold for any
orthonormal basis 
$\{ | \varphi_n \rangle \}_{n\in {\bf N}}$ of $\lt$:
\begin{eqnarray*}
\langle \, |\varphi_n \rangle \, ,\, |\varphi_m \rangle \, \rangle _{L^2}
\equiv
\langle \varphi_n  | \varphi_m \rangle & = & \delta_{nm}
\\
\sum_{n \in {\bf N}}
| \varphi_n \rangle \langle \varphi_n | & = & {\bf 1}_{L^2}
\ \ .
\end{eqnarray*}
Here, the last relation means that 
\begin{eqnarray*}
|\psi \rangle & = &
\sum_{n \in {\bf N}}
| \varphi_n \rangle \langle \varphi_n | \psi \rangle
\qquad  \ \mbox{for all} \ \  | \psi \rangle \in \lt
\\
\mbox{or} \quad
| \psi (x) \rangle & = &
\sum_{n \in {\bf N}}
| \varphi_n (x) \rangle \langle \varphi_n | \psi \rangle
\quad \mbox{for all} \ \  x \in \br
\ \ .
\end{eqnarray*}
In the same vein, the projector 
$P_{\psi}$ on 
$| \psi \rangle \in \lt$ 
can be written as 
$P_{\psi} = | \psi \rangle \langle \psi |$.
 
For operators, it is convenient to use the notation \cite{ct}
\[
| A \psi \rangle \equiv A | \psi \rangle
\ \ ,
\]
while avoiding the interpretation (\ref{cohe}) of matrix elements 
which represents a source of ambiguities; 
a matrix element can then be written 
in any of the following forms:
\[
\langle \varphi | A | \psi \rangle  =
\langle \varphi | A  \psi \rangle  =
\langle A^{\dag} \varphi | \psi \rangle =
\langle \psi | A^{\dag} \varphi \rangle ^{\ast} =
\langle \psi | A^{\dag} | \varphi \rangle ^{\ast}
\ \ .
\]
The insertion of the identity operator 
or of a projection operator 
are realized as in  
expressions (\ref{in1}) and (\ref{in2}), respectively.
 
Finally, the 
notation $|n..m \rangle$, instead of $\varphi_{n..m}\,$, for 
the vectors of a Hilbert space basis indexed by $n,..,m$
is quite useful for writing matrix elements\footnote{However, 
beware of the fact
that the resulting matrix represents a linear operator on the 
infinite dimensional Hilbert space $l_2$ so that 
one has to worry about its domain of definition:
the existence of matrix realizations and their mathematical pitfalls
are discussed in reference \cite{sg}.},  
\[
a_{n..m,n^{\prime} .. m^{\prime}} =
\langle n..m | A |
n^{\prime} .. m^{\prime} \rangle
\ \ .
\]
Hence, by allowing for some flexibility, 
one can benefit of the advantages of Dirac's notation 
while avoiding its inconveniences.

\chapter{Conclusion}
 
Let us try to draw some conclusions from the previous discussions, 
in particular for the teaching of quantum mechanics. 

Physics and mathematics are two different sciences and one can 
fully justify that a physicist's presentation does not 
take into account a perfect mathematical rigor even if the author 
completely masters this one. In physics, 
it probably is an art to use a minimum of mathematics while 
remaining precise enough in ones reasoning and  presentation
that a mathematical physicist can complete all technical details   
 without ambiguities and thereby establish the results and 
 their domain of validity in an irrefutable manner. 
For a quantum mechanics course, 
such an approach amounts to providing 
precise definitions in the beginning (for linear operators 
on $\lt$) while avoiding systematic discussions of  
mathematical details (domains of definition, distributions, ...)
in the sequel. 
On the other hand, any approach which starts with 
a symbolic calculus that 
is quite difficult to render rigorous (and thus capable  
of precise conclusions) seems questionable. 
This is all the more true since the first approach is not more complicated 
and since it is based on a standard, well developed 
mathematical theory finding applications in many other fields 
of physics (dynamical systems, relativity, optics, ...). 

\medskip 

For the reader's convenience, we conclude with a 
{\underline{\em short guide to the literature}}
which only considers the presentation of the 
mathematical formalism and {\em not} the one of the physical 
applications. In view of the vast number of textbooks 
on the subject, such a list can only be incomplete
and provide some hints for the interested reader. 

The physics books which do not follow the symbolic calculus approach 
(and which mention the domains of definition as well as the difference 
between Hermitian and self-adjoint operators) are not very numerous:
let us cite the monographs \cite{gap} 
which are not based on the abstract Hilbert space  
formalism and which make a
liberal use of Dirac's notation any time this 
appears to be beneficial
(see also \cite{bal, sche}). 
Reference \cite{mue} is a 
nice mathematically minded introduction 
to quantum mechanics which  does not 
strive for an overly complete  rigor.
A presentation  which is comparable, though more mathematical 
and oriented towards the conceptual foundations, is given in 
\cite{jmj} while the treatises
\cite{ber,th,rj,gl}
can be qualified as belonging to the field of mathematical physics. 
Among the textbooks \cite{ct}, those of Peebles, Schwabl and
Messiah  discuss 
in detail wave mechanics and its geometric structure before presenting 
the invariant formalism and Dirac's notation. 
Finally, we also mention some texts which avoid the 
invariant formalism and a rigid use of Dirac's notation, 
though they do not discuss the mathematical details concerning 
operators on 
$\lt$: apart from the `classics' \cite{ll}, these are the elementary and
modern introductions \cite{pd} which clearly present the principles of the
theory while applying a strict minimum of useful mathematics.

\vspace{14mm}
 
{\bf \Large Acknowledgments}
 
\vspace{5mm}
 
I seize the opportunity to thank the superb teachers from whom 
I could learn quantum mechanics in
G\"ottingen (H.~Goenner, H.~Roos, F.~Hund ($\dagger$1997),
J.~Yngvason, H.~Reeh) and in Berkeley (G.W.~Mackey).
Thanks to A.~Bilal for explaining a bit the invariant formalism 
to me upon my arrival in France! 
I express my gratitude to I.~Laktineh, M.~Kibler
and R.~Barbier for enduring my innumerable complaints, 
for all our discussions on quantum mechanics and for their 
comments on the text. 
Thanks to D.~Maison, P.~Breitenlohner, S.~Fleck, 
H.~Grosse, H.~K\"uhn,
R.~McDermott, B.F.~Samsonov and the very much 
regretted K.~Baumann ($\dagger$1998) 
for their pertinent remarks on the examples!
I would also like to express my gratitude to 
B\'en\'edicte Bruckert and Annabelle
Pontvianne for numerous discussions on physics and 
mathematics. The revised version of the text 
owes at lot to the unknown referees and to their 
constructive remarks on the original presentation.

\newpage
 
\appendix

\chapter{There is no surprise}
 
The solution of the problems and apparent contradictions 
encountered in section 2.1
can be rephrased in the following way \cite{rs}:
the theory of linear operators on infinite dimensional vector 
spaces is more complicated and interesting than the theory  
of finite order matrices. Let us now discuss   
the aforementioned problems while applying the mathematical results 
of section 2.2.

\bigskip
 
\underline{{\bf (1)}}
Suppose  the commutation relation 
$[P,Q] = \ds{\hbar \over \ri} \, {\bf 1}$ is satisfied by operators 
$P$ and $Q$ acting on a Hilbert space $\hs$ of finite dimension 
$n$ (i.e. $\hs \simeq {\bf C}^n$). 
In this case, $P$ and $Q$ 
can be realized by $n\times n$ matrices, the trace is a 
well defined operation and we obtain the result 
\[
0 = {\rm Tr} \, [P,Q ] 
\stackrel{(\ref{hei})}{=} {\rm Tr} \, ( \ds{\hbar \over \ri}
{\bf 1}_n ) =
\ds{\hbar \over \ri}\; n
\ \ .
\]
From this senseless result,
one concludes that Heisenberg's relation
cannot be realized on a {\em finite} dimensional Hilbert space. 
Thus, quantum mechanics has to be formulated on an
infinite dimensional Hilbert space: on such a space,  
the trace is not anymore a well defined operation for all operators 
(in particular, the trace of the operator ${\bf 1}$ does not 
exist) and therefore one can no more  deduce a contradiction from 
Heisenberg's commutation relation in the indicated manner. 

An inconsistency can still be deduced in another way 
on an infinite dimensional Hilbert space 
by assuming that 
$P$ and $Q$ are both bounded operators \cite{rs};
accordingly, 
at least one of the operators $P$ and $Q$ satisfying Heisenberg's
relation has to be unbounded and therefore this fundamental relation cannot
be discussed without worrying about the domains of definition of operators. 
 
\bigskip
 
\underline{{\bf (2)}}
As noted in section 2.2.2, 
the maximal domain of definition of the operator 
$P= \ds{\hbar \over \ri} \, \ds{d \over dx}$ 
on the Hilbert space $\lt$
is
\[
{\cal D}_{\rm max} (P) = \{ \psi \in \lt \, | \,
\psi^{\prime} \in \lt \}
\ \ .
\]
The functions belonging to ${\cal D}_{\rm max} (P)$ therefore enjoy  
certain regularity properties and their derivative is square 
integrable on $\br$. In particular, these functions are continuous and 
their limit for $x \to \pm \infty$ 
is zero \cite{ri, bgc}; this implies that the operator $P$,
acting on ${\cal D}_{\rm max} (P)$, is Hermitian.
The aforementioned 
function, which is unbounded at infinity,  
is differentiable,
but its derivative is not square integrable and therefore 
 it does not belong to 
${\cal D}_{{\rm max}}(P)$.

Another choice of domain for $P$ is given by the 
Schwartz space ${\cal S}(\br)
\subset {\cal D}_{{\rm max}} (P)$.
In this case, the functions on which the operator $P$ acts,  
 do even have a rapid decrease at infinity. 
 
\bigskip
 
\underline{{\bf (3a)}}
The Schwartz space $\st \subset \lt$ is an invariant  
domain of definition for the operators $P$ and $Q$ 
and thereby also for  
$A= PQ^3 +Q^3 P$ :
\[
A : \st \longrightarrow \st
\ \ .
\]
Integration by parts shows that the so-defined operator 
$A$ is Hermitian :
\[
\langle g , A f \rangle  =
\langle A g , f \rangle
\qquad \mbox{for all} \ f,g \in {\cal D}(A) = \st
\ \ .
\]
The function $f$ given by (\ref{fon}) belongs to the Hilbert space 
$\lt$, but it does not belong to the domain 
of $A$, since it does {\em not} 
decrease more rapidly than the inverse 
of any polynomial at infinity: for instance, 
$x^3 f(x) \propto  x^{3/2} \, {\rm exp}\, [-1 /(4x^2)]$ 
is not bounded for 
$x \to +\infty$.
 By way of consequence, 
$\hbar /\ri$ is {\em not} an eigenvalue of $A$.
 
However \cite{blt}, 
$\hbar /\ri$ is an eigenvalue of the operator  
$A^{\dag}$ which has the same operating prescription as $A$.  
Before discussing this point, it is preferable to first consider the
solution of the other problems.  
 
\bigskip
 
\underline{{\bf (4a)}}
The astonishing 
results we referred to in this example, 
indicate that it is not enough 
to verify that an operator is Hermitian to identify it with  
an observable: 
we already dealt with this well-known fact 
in section 2.2.2 where 
we showed that the operator $P$ 
and its adjoint act in the same way, but admit  
different domains of definition 
(given by (\ref{dir}) and (\ref{dadj}), respectively).
Thus, we established that $P$ 
is Hermitian, but not self-adjoint.
 
The results concerning the spectrum of $P$ 
which are cited in this example, indicate that 
the spectrum of a Hermitian operator is not simply the set 
of its proper or generalized 
eigenvalues.
Since this point is a more subtle one, we postpone 
its discussion.

\bigskip
 
\underline{{\bf (5)}}
The operator of multiplication by $\varphi$ on the Hilbert space 
$\hs = L^2
([0,2 \pi ], d\varphi )$ 
is everywhere defined and self-adjoint: 
\[
\langle g , \varphi \, f \rangle
 = \langle  \varphi \, g , f \rangle \qquad \mbox{for all} \ \
g,f \in \hs
\ \ .
\]
The discussion in section 2.2.2
concerning the operator $P = \ds{\hbar \over \ri} \,
\ds{d \over dx}$ on 
$L^2
([0,1], dx )$ applies verbatim  to the operator 
$L_z = \ds{\hbar \over \ri} \,
\ds{d \over d\varphi}$ on 
$L^2
([0,2 \pi ], d\varphi )$ : integration by parts yields
\begin{equation}
\label{uti}
\int_0^{2\pi} \! d \varphi \, (  \overline{g} \, L_z f
-
\left( \overline{\ds{\hbar \over \ri} \,
\ds{dg \over d\varphi} }  \right)  f ) (\varphi )
= \ds{\hbar \over \ri} \left[ \overline{g(2 \pi)} f (2 \pi) -
\overline{g( 0)} f (0) \right]
\quad \mbox{for all} \ \, f \in {\cal D}(L_z)
\ \ .
\end{equation}
Due to the periodic character of the polar angle, 
the functions belonging to the domain of $L_z$
are periodic\footnote{Concerning this point, we note that 
the use of polar coordinates assigns a distinguished role to 
the polar semi-axis $\varphi =0$, while this axis is not privileged 
if one chooses other coordinate systems like Cartesian 
coordinates: 
therefore a discontinuity of wave functions on this axis 
($f(2 \pi)= {\rm e}^{\ri \alpha}
f(0)$ with 
$\alpha \neq 0$)
has no raison d'\^etre.}:
\begin{equation}
\label{oaa}
L_z = \ds{\hbar \over \ri} \, \ds{d \over d\varphi} \quad , \quad
{\cal D} (L_z ) = \{ f \in \hs \; | \;
f^{\prime} \in \hs \ \; {\rm and} \ \; f(0) = f(2 \pi ) \}
\ \ .
\end{equation}
Accordingly, the surface term in (\ref{uti}) vanishes
if and only if $g(0) = g(2 \pi)$ :
this implies that $L_z^{\dag}$ operates in the same way as $L_z$
and that it admits the same domain, hence 
the operator (\ref{oaa}) is self-adjoint. 

In order to determine the domain of definition of the commutator 
$[L_z, \varphi ]$, we note that for any two operators $A$ and $B$,
we have 
\begin{eqnarray}
{\cal D} (A+B) & = & {\cal D}(A) \cap {\cal D}(B)
\\
{\cal D} (AB) & = & \{ f \in {\cal D}(B)\, | \, Bf \in {\cal D}(A) \}
\nonumber
\ \ .
\end{eqnarray}
 Thus, 
${\cal D} ([L_z, \varphi]) = {\cal D}(L_z \varphi ) \cap
{\cal D} (\varphi L_z )$ with 
\begin{eqnarray*}
{\cal D} (\varphi L_z )
& = & \{ f \in {\cal D}(L_z) \, | \, L_z f \in {\cal D}(\varphi )= \hs \}
\ = \ {\cal D} (L_z )
\\
{\cal D} (L_z  \varphi) & = &
\{ f \in {\cal D}(\varphi) =\hs \, | \, \varphi  f \in {\cal D}(L_z) \}
\ \ .
\end{eqnarray*}
But the function 
$\tilde f \equiv \varphi f$, which appears in the last expression,  
takes the values 
\begin{eqnarray*}
\tilde f (0) & = & (\varphi f)(0) \ = \ 0
\\
\tilde f (2 \pi ) & = & (\varphi f)(2\pi ) \ = \ 2 \pi \, f (2 \pi)
\end{eqnarray*}
 and $\tilde f \in {\cal D}(L_z)$ implies 
$\tilde f (0) =
\tilde f (2 \pi )$, i.e. $f(2\pi ) = 0$.
 
In summary, 
\begin{eqnarray}
{\cal D} (\varphi L_z )
& = & {\cal D} (L_z )
\nonumber
\\
{\cal D} (L_z  \varphi) & = &
\{ f \in \hs \, | \, f^{\prime} \in \hs \ \ {\rm and} \ \
f (2 \pi ) =0 \}
\\
{\cal D} ([ L_z , \varphi ]) & = &
\{ f \in \hs \, | \, f^{\prime} \in \hs \ \ {\rm and} \ \
f (0) =0 =
f (2 \pi ) \}
\ \ .
\nonumber
\end{eqnarray}
The eigenfunctions 
$\psi_m ( \varphi ) = \ds{1 \over \sqrt{2\pi} } \, {\rm exp}\,
( \ri  m \varphi )$ of $L_z$
do not belong to the domain of definition of 
$[ L_z , \varphi ]$ since they do not vanish at the points 
$0$ and 
$2\pi$ : therefore the 
derivation (\ref{deriv}) does not make any sense. 
 
\bigskip
 
\underline{{\bf (6)}}
Let us consider two observables $A$ and $B$ 
(i.e. self-adjoint operators 
on the Hilbert space $\hs$) and a state $\psi$ (i.e. a unit vector 
belonging to $\hs$).
The uncertainty relation for $A,B$ is usually written in the form 
\cite{gap}
\begin{equation}
\label{ri}
\Delta_{\psi} A \cdot
\Delta_{\psi} B \geq \ds{1 \over 2} \, \mid \langle \psi ,
\ri [A, B] \psi \rangle \mid
\ \ ,
\end{equation}
where 
$(\Delta_{\psi} A)^2 = \| \left( A -
\langle A\rangle_{\psi} {\bf 1} \right) \psi \|^2$
with 
$\langle A\rangle_{\psi} = \langle \psi , A\psi \rangle$ and likewise 
for $B$.
Thus, the left-hand side of relation (\ref{ri}) is defined for 
$\psi \in {\cal D}(A) \cap {\cal D}(B)$
(which is precisely the subspace of $\hs$ containing all states 
$\psi$ for which the uncertainties 
$\Delta_{\psi} A$ and 
$\Delta_{\psi} B$ both have a physical meaning). 
On the other hand, the right-hand side is only defined on the subspace 
${\cal D} ([A,B]) =
{\cal D} (AB) \cap
{\cal D} (BA)$ which is much smaller in general. 

However, $A,B$ being self-adjoint, relation (\ref{ri}) can be rewritten 
in the form \cite{krau}
\begin{equation}
\label{rir}
\Delta_{\psi} A \cdot
\Delta_{\psi} B \geq \ds{1 \over 2} \, \mid \ri \langle A\psi ,
B \psi \rangle - \ri \langle B\psi , A \psi \rangle \mid
\ \ ,
\end{equation}
where the domain  of the right-hand side 
now coincides with the one of the left-hand side, i.e.  
${\cal D}(A) \cap {\cal D}(B)$.
Thus, the product of  
uncertainties for two observables 
$A$ and $B$ is not determined by their commutator, but by the 
 Hermitian sesquilinear form\footnote{i.e. 
 $\Phi_{A,B} (f,g)$ is linear with respect to $g$, 
 antilinear with respect to $f$
 and $\Phi_{A,B} (g,f) = \overline{\Phi_{A,B} (f,g)}$.}
\[
\Phi_{A,B} (f,g) =
\ri \langle A f, Bg \rangle
-\ri \langle B f , A g \rangle
\qquad \mbox{for all} \ \,  f,g \in
{\cal D}(A) \cap {\cal D}(B)
\ \ .
\]
The derivation of inequality (\ref{rir}) is the same as the 
derivation 
of (\ref{ri}) (see for instance \cite{krey} for the latter). 
It can be done in a few lines: 
 let $\psi \in {\cal D}(A) \cap {\cal D} (B)$ and let 
\[
\hat A = A - \langle A \rangle _{\psi} {\bf 1}
\qquad , \qquad
\hat B = B - \langle B \rangle _{\psi} {\bf 1}
\ \ ;
\]
by using the fact that $A$ and $B$ are self-adjoint and by applying
the triangle as well the Cauchy-Schwarz inequalities,
we obtain the inequality (\ref{rir}) :
\begin{eqnarray*}
\mid \ri \langle A \psi , B \psi  \rangle
-\ri \langle B \psi , A \psi \rangle \mid
& = &
\mid \ri \langle \hat A \psi , \hat B \psi  \rangle
-\ri \langle \hat B \psi , \hat A \psi \rangle \mid
\\
& \leq &
\mid  \langle \hat A \psi , \hat B \psi  \rangle  \mid +
\mid  \langle \hat B \psi , \hat A \psi  \rangle  \mid
\ = \ 2 \,
\mid  \langle \hat A \psi , \hat B \psi  \rangle  \mid
\\
& \leq &  2 \,
\| \hat A \psi \| \cdot
\| \hat B \psi \|  \ = \
2\, \Delta_{\psi} A \cdot
\Delta_{\psi} B
\ \ .
\end{eqnarray*}

Let us now show that this modification is not simply a cosmetic one. 
For 
$A=P= \ds{\hbar \over \ri} \, \ds{d \over dx}$ and $B=Q=x$
 on $\hs = L^2 ( {\bf R}, dx)$, the right-hand side of inequality 
(\ref{rir}) is easily evaluated using integration by parts;  
it implies the well-known uncertainty relation 
$\Delta_{\psi} P \cdot
\Delta_{\psi} Q \geq \ds{\hbar  \over 2}$ for  $\psi \in
{\cal D}(P) \cap {\cal D}(Q)$.
On the other hand, for 
$A= L_z = \ds{\hbar \over \ri} \, \ds{d \over
d\varphi}$ and $B=\varphi$
 on $\hs = L^2 ( [0,2\pi ] , d\varphi )$, the surface term 
 occurring upon integration by parts does not vanish and it leads
 to the uncertainty relation 
\begin{equation}
\label{lphi}
\Delta_{\psi} L_z \cdot
\Delta_{\psi} \varphi \geq \ds{\hbar  \over 2}
\ \mid 1 - 2 \pi | \psi (2 \pi ) | ^2 \mid \qquad
\mbox{for all} \ \; \psi \in {\cal D}(L_z) \cap {\cal D}(\varphi)
= {\cal D} (L_z)
\ .
\end{equation}
Thus, the product of the uncertainties 
$\Delta_{\psi} L_z$ and 
$\Delta_{\psi} \varphi$ 
may become smaller than 
$\hbar /2$
- see Galindo and Pascual \cite{gap} for an example. 
For states   
$\psi$ which belong to ${\cal D} ([L_z , \varphi ] )$, i.e. 
satisfy 
$\psi (2 \pi )=0$, the inequality  (\ref{ri}) 
can also be applied and it yields the same result as (\ref{lphi}).
 
While the inequality (\ref{lphi}) is mathematically correct, 
it is not acceptable in the present 
form from the point of view of physics: 
if one defines the average value and the uncertainty of the observable 
$\varphi$ by the usual formulae, then 
these expressions do not have the 
adequate transformation properties with respect to rotations,
for which 
\[
\psi (\varphi ) \to 
\left( {\rm e} ^{ {\ri \over \hbar} \alpha L_z }  \psi \right) 
(\varphi )
= \psi (\varphi + \alpha)
\ \ .
\]
We refer to the literature 
\cite{ju,krau}
for a slight modification of (\ref{lphi})
 which takes this problem into account, as well for 
estimates of the product  
$\Delta_{\psi} L_z \cdot
\Delta_{\psi} \varphi$ 
which do not explicitly depend on the particular state 
$\psi$ that one considers. 
Similar issues concerning the phase and number operators 
in quantum optics are discussed in reference \cite{car}.

\bigskip
 
\underline{{\bf (7)}}
A purely formal solution of the problem can be obtained,
if one considers the wave function to be defined 
on the entire real axis, rather than
limiting oneself to the interval $[-a, +a]$.
In fact, for the function $\psi$ defined by (\ref{psi}),
the discontinuity of 
$\psi^{\prime \prime}$ at the points $x = \pm a$ implies that 
$\psi^{\prime \prime \prime \prime}$ is given by derivatives 
of Dirac's generalized function: 
\[
\psi^{\prime \prime \prime \prime} (x) = -
{\sqrt{15} \over 2 a^{5/2}} \left[
\delta^{\prime} (x+a) - \delta^{\prime} (x-a) \right]
\qquad {\rm for} \ \, x \in \br
\ \ .
\]
Substitution of this expression in $\langle \psi , H^2 \psi \rangle$
then leads to the same non-vanishing result as the  
evaluation of $\sum_{n=1}^{\infty} E_n^2 p_n$.
The mentioned inconsistency therefore originates from the fact that 
one has not properly taken into account the boundary conditions 
in the calculation (\ref{caf}).

In the sequel, we will show 
how a rigorous reasoning limited to the interval 
$[-a , +a]$ allows to incorporate the boundary conditions 
and to confirm the non-vanishing result for 
$\langle H^2 \rangle_{\psi}$.
To start with, we define 
$H$ and $H^2$ as self-adjoint operators on the Hilbert space 
$\hs = L^2 ([-a,+a], dx)$.
 
The infinite potential well is a mathematical idealization 
which is to be interpreted as the limit $V_0 \rightarrow \infty$
of a potential well of finite height $V_0$. For the latter, 
one finds that, outside the well, the wave functions 
of stationary states tend to zero if 
$V_0 \rightarrow \infty$; by way of consequence, $\psi(\pm a)=0$
is the appropriate boundary condition for the particle 
confined to the inside of 
the infinite well. Let us now study whether $H$ is self-adjoint,
when acting on sufficiently smooth  functions 
satisfying $\psi (\pm a)=0$ : by virtue of two integration 
by parts, we find  
\begin{eqnarray*}
\langle \varphi , H \psi \rangle & \equiv &
\ds{- \hbar^2 \over 2m} \int_{-a}^{+a} dx \, \overline{\varphi(x)}
\psi^{\prime \prime}(x)
\\
&= &
\ds{- \hbar^2 \over 2m} \int_{-a}^{+a} dx \, \overline{\varphi^{\prime
\prime}(x)}
\psi (x) +
\ds{\hbar^2 \over 2m}
\left[ (\overline{\varphi^{\prime}} \psi - \overline{\varphi}
\psi^{\prime} )(x) \right]_{-a}^{+a}
\\
& = & \langle H^{\dag} \varphi , \psi \rangle -
\ds{\hbar^2 \over 2m}
\left[ \overline{\varphi} (a)\psi^{\prime} (a) - \overline{\varphi}(-a)
\psi^{\prime} (-a) \right]
\ \ .
\end{eqnarray*}
Since we do not have any constraints on $\psi^{\prime}(\pm a)$,
the surface term vanishes if and only if $\varphi (\pm a)=0$.
In summary, $H^{\dag}$ operates in the same way as $H$
and the functions $\varphi$ belonging to its domain of definition 
satisfy the same conditions as those belonging to the domain of $H$. 
Hence, the operator 
$H= \ds{-\hbar^2 \over 2m} \, \ds{d^2 \over dx^2}$ acting on 
$\hs$, with the domain of definition 
\begin{equation}
{\cal D} (H) = \{ \psi \in \hs \; \mid \; \psi^{\prime \prime}
\in \hs \ \; {\rm and} \ \; \psi(\pm a) =0 \}
\end{equation}
represents a self-adjoint operator (i.e. an observable). 
Its spectrum, which has been made explicit in equation (\ref{pui}),
is discrete and non-degenerate and the associated eigenfunctions 
are given by 
\[
\varphi_n (x) = \left\{
\begin{array}{ll}
\ds{1 \over \sqrt a} \  {\rm sin}  \left( \ds{n\pi \over 2a} x \right)
& \quad {\rm for} \ \, n=2,4,6,...
\\
\ds{1 \over \sqrt a} \  {\rm cos}  \left( \ds{n\pi \over 2a} x \right)
& \quad {\rm for} \ \, n=1,3,5,... \ \ .
\end{array}
\right.
\]
Accordingly, the spectral decomposition of $H$ reads
$H= \sum_{n=1}^{\infty} E_n {\cal P}_n$ where ${\cal P}_n$
 denotes the projector on the normed state $\varphi_n$ :
${\cal P}_n \psi = \langle \varphi_n , \psi \rangle \, \varphi_n$.
 
By virtue of the spectral theorem \cite{rs},
the operator $H^2$ is defined in terms of the spectral 
decomposition of  $H$,
\begin{equation}
\label{h2}
H^2 = \sum_{n=1}^{\infty} E_n^2 \, {\cal P}_n
\ \ ,
\end{equation}
which implies that 
$H^2 \varphi_n = E_n^2 \varphi_n$.
In order to determine explicitly the domain of definition on which this
operator 
is self-adjoint, one successively performs four integrations by parts: 
\begin{eqnarray*}
\langle \varphi , H^2 \psi \rangle & \equiv &
\ds{\hbar^4 \over 4m^2} \int_{-a}^{+a} dx \, \overline{\varphi(x)}
\psi^{\prime \prime \prime \prime}(x)
\\
&= &
\ds{\hbar^4 \over 4m^2} \int_{-a}^{+a} dx \, \overline{\varphi^{\prime
\prime \prime \prime}(x)}
\psi (x) +
\ds{\hbar^4 \over 4m^2} \left[
( \overline{\varphi} \psi^{\prime \prime \prime} - \overline{\varphi
^{\prime}} \psi^{\prime \prime} +
\overline{\varphi^{\prime \prime}} \psi^{\prime} -
\overline{\varphi^{\prime \prime\prime}} \psi )(x)
\right]_{-a}^{+a}
.
\end{eqnarray*}
The boundary conditions $\psi (\pm a) =0 = \varphi (\pm a)$
of the infinite well eliminate the first and the last contribution of the
surface term. In order to annihilate the others, there are 
different possibilities, e.g. 
$\psi^{\prime} (\pm a) =0 = \varphi^{\prime} (\pm a)$ or
$\psi^{\prime \prime} (\pm a) =0 = \varphi^{\prime \prime} (\pm a)$.
But, according to the definition (\ref{h2}) of $H^2$,
the eigenfunctions $\varphi_n$ of $H$ must belong to the domain 
of $H^2$: since these functions satisfy 
$\varphi_n ^{\prime \prime}
(\pm a)=0$, the  domain of definition of the observable $H^2$ is 
\begin{equation}
\label{dh2}
{\cal D} (H^2) = \{ \psi \in \hs \; \mid \; \psi^{\prime \prime
\prime \prime}
\in \hs \ \; {\rm and} \ \; \psi(\pm a) =0 = \psi^{\prime \prime}
(\pm a) \}
\ \ .
\end{equation}
 
 We note that this represents but one way to render the operator 
$\ds{\hbar^4 \over 4m^2} \, \ds{d^4 \over dx^4}$ 
on $L^2 ([-a,+a])$
self-adjoint among many other possibilities 
(determined by other boundary conditions,
e.g.  $\psi (\pm a ) = 0 = \psi^{\prime} (\pm a)$); 
but it is the one corresponding to the   
physical system that we consider here. 

Let us now come to the paradox pointed out in our example. For 
$\psi \in
{\cal D}(H^2) \subset {\cal D}(H)$,
 the decomposition (\ref{h2})
yields 
\begin{eqnarray}
\nonumber
\langle H^2 \rangle_{\psi} \equiv 
\langle \psi , H^2 \psi \rangle 
& \stackrel{(\ref{h2})}{=} & 
\left\langle \psi ,
\sum_{n=1}^{\infty} E_n^2 \, {\cal P}_n \psi \right\rangle 
\ = \ 
\sum_{n=1}^{\infty} E_n^2 \langle \psi , {\cal P}_n \psi \rangle
\\
&= &
\sum_{n=1}^{\infty} E_n^2 \, p_n  
\end{eqnarray}
 with $p_n = |  \langle \varphi_n , \psi \rangle  |^2$.
 If $\psi \in {\cal D} (H)$, we can reach the same result 
 in another way by using the fact that the projectors 
${\cal P}_n$ are self-adjoint and orthogonal 
(i.e. ${\cal P}_n {\cal P}_m = \delta_{nm} {\cal P}_n$) :
\begin{eqnarray}
\| H \psi \|^2 
&\equiv & 
\langle H \psi ,  H\psi \rangle
\ = \ \left\langle \sum_{n=1}^{\infty} E_n {\cal P}_n \psi ,
\sum_{m=1}^{\infty} E_m {\cal P}_m \psi \right\rangle 
\ = \ 
\sum_{n,m=1}^{\infty} E_n E_m \langle {\cal P}_n \psi , {\cal P}_m \psi
\rangle
\nonumber
\\
& = &
\sum_{n,m=1}^{\infty} E_n E_m \langle \psi , {\cal P}_n {\cal P}_m \psi
\rangle
\nonumber
\\
&=&
\sum_{n=1}^{\infty} E_n^2 \, p_n 
\ \ .
\end{eqnarray}
 The function
$\psi(x) = \sqrt{15}/(4a^{5/2}) \,
(a^2 - x^2)$ of our example does not satisfy  
$\psi^{\prime \prime} (\pm a) =0$, hence it does not belong 
to the domain of definition of $H^2$: thus the expression 
$\langle \psi, H^2 \psi \rangle$ is not defined, because the  
 variable $H^2$ that it involves is not simply characterized  by  
its operating prescription, but also by its domain of definition. 
(In other words: although the integral in equation
(\ref{caf}) is properly evaluated, it cannot be
identified with  $\langle \psi , H^2 \psi \rangle = \langle H^2
\rangle_{\psi}$ for the function $\psi$
considered here.)
On the other hand, we have $\psi \in {\cal D} (H)$ and the average value 
$\langle H^2 \rangle _{\psi}$
can be evaluated according to 
$\sum_{n=1}^{\infty} E_n^2 \, p_n$ 
or, equivalently, according to 
\[
\| H \psi \|^2 \equiv 
\int_{-a}^{+a} dx \, | (H\psi )(x) | ^2 =
\ds{\hbar^4 \over 4m^2}
\int_{-a}^{+a} dx \, | \psi^{\prime \prime} (x) | ^2 =
\ds{15 \hbar^4 \over 8m^2 a^4}
\ \ .
\]

\bigskip
 
\underline{{\bf (4b)}}
Let us now consider the second point mentioned in example 4, 
i.e. the spectrum of $P$.
As indicated in section 2.2.2, 
the spectrum of an 
operator $P$, which is not self-adjoint, 
contains in general
a part called {\em residual spectrum}:
by definition, these are all numbers $z \in {\bf C}$
which are not eigenvalues of $P$, but for which $\bar z$
is an eigenvalue of $P^{\dag}$.
In the present example, the domains of $P$ and $P^{\dag}$
are given by 
(\ref{dir}) and (\ref{dadj}), respectively;
thus the functions belonging to ${\cal D}(P^{\dag})$
do not satisfy any boundary conditions, while
those belonging to ${\cal D}(P)$ vanish at the boundaries
$x=0$ and $x=1$. 
Since the functions 
$\varphi_p (x)=  \, {\rm exp} \, ( \ds{\ri \over \hbar} px )$
 with $p \in \bc$ are solutions
of the eigenvalue equation for 
$P^{\dag}$,
\begin{equation}
\label{spr}
(P^{\dag} \varphi_p)(x) = p \,  \varphi_p(x)
\qquad ( \varphi_p \in {\cal D}(P^{\dag})
\ , \ \varphi_p \not \equiv 0 )
\ \ ,
\end{equation}
all complex numbers are eigenvalues of $P^{\dag}$.
But none of these is an eigenvalue of $P$, because 
$\varphi_p$ does not vanish on the boundaries and therefore
does not belong to ${\cal D}(P)$. 
By way of consequence, the residual spectrum of $P$ is $\bc$. 
In fact, this represents the 
 complete spectrum of $P$ since its  
discrete and continuous spectra
are empty. 

This conclusion resolves the puzzle of example 4, 
but it leaves the physicist perplexed: 
since $P$ is not 
 self-adjoint, its spectrum does not admit a direct 
 physical interpretation. However, we will show right away 
 that it contains information which is important for physics. 

In order to study whether the domain of the Hermitian 
operator $P$ 
can be enlarged in such a way that $P$ becomes self-adjoint, 
it is suitable to apply {\em von Neumann's theory of 
defect  indices} \cite{rs, sg}: 
according to this theory, 
one has to study the complex eigenvalues of 
$P^{\dag}$ (and this is precisely what we just did). 
As a special case of (\ref{spr}), we have\footnote{For 
the present mathematical considerations, one 
ignores the dimensions of the eigenvalues which should coincide
with the dimensions of the operators.} 
\[
P^{\dag} \varphi_{\pm} = \pm \ri \, \varphi_{\pm}
\qquad {\rm with} \ \ \varphi_{\pm} (x) = {\rm e}^{\mp x / \hbar}
\ \ ,
\]
or  
$(P^{\dag} \mp \ri {\bf 1})  \varphi_{\pm} =0$.
Thus, the kernel of the operator $P^{\dag} \pm \ri {\bf 1}$ 
is a one-dimensional vector space: 
\begin{eqnarray}
\label{defe1}
n_- (P) & \equiv & {\rm dim \; Ker} \,
(P^{\dag} +  \ri {\bf 1}) =1
\nonumber
\\
n_+ (P) & \equiv & {\rm dim \; Ker} \,
(P^{\dag} -  \ri {\bf 1}) = 1
\ \ .
\end{eqnarray}
The natural numbers 
$n_+ (P)$ and 
$n_- (P)$ are called the {\em deficiency} or {\em defect indices of} $P$. 
Their usefulness is exhibited by the following result:
\begin{theo}[Criterion for self-adjointness]
Let $A$ be a Hermitian operator with deficiency indices
$n_+$ and $n_-$. 

\noindent 
{\rm (i)} 
$A$ is self-adjoint if and only if $n_+ = 0= n_-$.
In this case (and only in this one), the spectrum of $A$ is 
a subset of the real axis. 

\noindent 
{\rm (ii)}
$A$ admits self-adjoint extensions (i.e. it is possible to 
render $A$ self-adjoint by enlarging its domain of definition)
if and only if $n_+ =  n_-$.
 If $n_+ >0$ and $n_- >0$, the spectrum of $A$ is the entire 
 complex plane. 

\noindent 
{\rm (iii)} 
If either $n_+ =0 \neq n_-$ or $n_- =0 \neq  n_+$,
the operator $A$ has no nontrivial self-adjoint extension.
Then, the spectrum of $A$ is, respectively, 
the closed upper, or the closed lower, complex half-plane.
\end{theo}
In the case $(ii)$, there exist explicit expressions 
for the possible self-adjoint extensions of $A$ \cite{rs}. 

In our example, we have $n_+ =  n_- >0$ ; therefore the operator $P$ 
is not
self-adjoint and its spectrum is the entire complex plane 
(what we are already aware of). 
The explicit expressions for the possible 
self-adjoint extensions of $P$ that we just 
alluded to,  are as follows \cite{rs}: 
for every real number  $\alpha$, the operator  
\begin{equation}
\label{alpha}
P_{\alpha} = \ds{\hbar \over \ri} \, \ds{d \over dx} \quad , \quad
{\cal D} (P_{\alpha}) = \{ \psi \in \hs \, | \,
\psi^{\prime} \in \hs \ \; {\rm and} \ \; \psi(0) =
{\rm e}^{\ri \alpha} \psi(1) \}
\end{equation}
is self-adjoint (as may easily be verified) 
and one has 
${\rm Sp} \, P_{\alpha} = \br$.

From the point of view of physics, the boundary condition 
$\psi(0) = 
{\rm e}^{\ri \alpha} \psi(1)$ means that everything that 
leaves the interval $[ 0,1]$ on the right-hand side, again enters 
the interval on the left-hand side with a certain phase-shift
(determined by $\alpha \in \br$):
this allows for the existence of states with a well-defined value of
momentum, whereas the boundary condition $\psi(0) = 0 = \psi(1)$
excludes such states. 
For $\alpha =0$, one has periodic wave functions and one 
recovers the self-adjoint operator (\ref{oaa}).

\bigskip
 
\underline{{\bf (3b)}}
We now come back to the statement that $\hbar /\ri$ is an eigenvalue of 
$A^{\dag}$.
For $f \in {\cal D}(A) = \st$, 
integration by parts yields
\begin{equation}
\label{surf}
\langle g, A f \rangle \; = \;  \langle \ds{\hbar \over \ri} \left[
(x^3 g)^{\prime} \; + \; 
x^3 g^{\prime} \right] , f \rangle +
2 \, \ds{\hbar \over \ri} \left[
x^3 (\bar g  f)(x) \right]_{-\infty}^{+\infty}
\ \ .
\end{equation}
Since $f$ decreases rapidly at infinity, 
the surface term on the right-hand side vanishes if the function $g$
does not grow faster than a polynomial at infinity. 
Assuming $g$ has this property, 
the previous equation implies that the operator 
$A^{\dag}$ acts in the same way as $A$,
\begin{equation}
\label{exa}
A^{\dag} g = \ds{\hbar \over \ri} \, \left[ (x^3 g)^{\prime}
+x^3 g^{\prime} \right]
= \ds{\hbar \over \ri} \, \left[ 3x^2 g
+ 2 x^3 g^{\prime} \right]
\ \ ,
\end{equation}
though its domain is larger than $\st$:
this domain contains all functions $g$ which are such that  
expression  (\ref{exa}) exists and is square integrable. 
(For all of these functions, the surface term in equation 
(\ref{surf}) vanishes automatically.)
 
In summary, the domain of definition of $A^{\dag}$ is larger 
than the one of $A$, thus the operator $A$ is not self-adjoint. 
Moreover, the function (\ref{fon}) does not belong to ${\cal D}(A)$,
but it belongs to 
${\cal D}
(A^{\dag})$, so $\hbar / \ri$ is an eigenvalue of $A^{\dag}$.
 
 To conclude, we briefly investigate whether    
 the domain of $A$ can be enlarged 
 so as to render $A$ self-adjoint. 
 For this problem, we again resort to von Neumann's theory. 
One easily checks that 
\[
A^{\dag} g_{\pm} = \pm \ri \, g_{\pm}
\qquad {\rm with} \ \
\left\{
\begin{array}{l}
g_{\pm} (x) =
|x|^{-3/2} \; {\rm exp} \, \left( \pm \ds{1 \over 4 \hbar x^2}
\right)  \quad {\rm for} \ x \neq 0 \\
g_- (0) = 0 \  \ .
\end{array}
\right.
\]
We have 
$g_- \in {\cal D} (A^{\dag} )$, but 
$g_+ \not\in {\cal D} (A^{\dag} )$ (due to the exponential growth of  
$g_+$ at the origin), hence 
\begin{eqnarray*}
n_- (A) & \equiv & {\rm dim \; Ker} \,
(A^{\dag} + \ri {\bf 1}) =1
\\
n_+ (A) & \equiv & {\rm dim \, Ker} \,
(A^{\dag} - \ri {\bf 1}) = 0
\ \ .
\end{eqnarray*}
From point $(iii)$ of the last theorem,
it now follows that there is no way to render the Hermitian
operator $A$ self-adjoint.

\medskip 
 
 Although the introduction of the residual spectrum looks,
 at first sight, like a 
 gratuitous and unphysical complication, the last two examples
 show that it is quite interesting from the point of view of  physics.
 In fact, 
the deficiency indices of $A$ (which are closely
 related to the residual spectrum of $A$) 
allow us to tell whether a given Hermitian operator 
can be rendered self-adjoint; and for those $A$ which can,
von Neumann's theory also  
provides a simple and
 constructive method  for determining 
all possible self-adjoint extensions,  
i.e. it explicitly describes all possible ways of turning 
 a Hermitian operator  into an observable. 
 
 A more intuitive understanding of the last two examples 
 can be obtained by looking at the potential eigenfunctions 
 of the involved operators and at their admissibility  
 for the physical problem under consideration. 
 For the momentum operator $P$ on the interval $[0,1]$,
 the plane wave ${\rm exp} \, (\ds{\ri \over \hbar} p x)$ 
 (with $p \in \br$) formally solves the eigenvalue equation
 for $P$, but it is  
 not compatible with the boundary conditions 
 $\psi (0) = 0 = \psi (1)$ of the infinite potential well; 
 on the other hand, the function
 $f_{\lambda} (x) \propto |x|^{-3/2} \, 
 {\rm exp} \, ( \ds{- \ri \lambda \over 4 \hbar x^2})$ 
can formally  be associated 
to the real eigenvalue $\lambda$ of 
 $A = PQ^3 +Q^3 P$, but it is not square integrable due to its singular
 behavior  at the origin. 
 Thus, the crucial constraints for turning Hermitian operators 
 into observables 
 come, respectively,  from the 
 boundary conditions for a problem on a compact (or semi-infinite)
interval and from the 
 condition of square integrability for a problem 
on the whole space\footnote{Actually, the lack of square 
integrability only points towards {\em potential} problems 
for self-adjointness and, to be more precise, 
we should say the following. The eigenfunctions 
associated with the continuous spectrum do not belong 
to the Hilbert space and thus are not square integrable
(section 2.2.3). However, in the case of differential operators,
these eigenfunctions enjoy certain smoothness properties 
(which imply some local square integrability) and what matters
for the continuous spectrum,  
is that the singularities of the functions
are ``mild enough''. For instance,
for $n =1,2,..$, the potential eigenfunctions $f^{(n)}_{\lambda}$
of $A_n =PQ^n + Q^nP$
satisfy   $|f^{(n)}_{\lambda} (x)| \propto |x|^{-n/2}$
and, for  $n \geq 3$, 
they are too singular at $x=0$ to allow for the self-adjointness
of $A_n$; however, $A_1$ and $A_2$ are self-adjoint,
when defined on appropriate domains.}.    
 (As a matter of fact, these are the very same conditions which lead
  to the quantization of energy levels, respectively,  
 on a finite interval 
 and on the whole space.)

\end{document}